\documentclass[traditabstract]{aa} 
\usepackage{orcidlink}
\usepackage{natbib}
\usepackage{float}
\usepackage{graphicx}
\usepackage{mwe}
\usepackage{txfonts}
\usepackage{hyperref}
\hypersetup{
     colorlinks   = true,
     citecolor    = blue
}

\usepackage{placeins}
\usepackage{siunitx}
\usepackage{gensymb}
\usepackage{caption,subcaption}
\usepackage{subcaption}
\usepackage{pgffor}
\usepackage{dblfloatfix}    
\usepackage{lscape} 
\usepackage{multirow}
\usepackage[export]{adjustbox}
\usepackage{numprint}
\npthousandsep{\,}

\def \msun{$\,M_{\odot}\,$}
\def \lsun{$\,L_{\odot}\,$} 
\def \mstar{$\,M_{\rm star}\,$}
\def \chisq{$\chi^2$}
\def \magperarcsec{mag arcsec$^{-2}$}

\def \Re{$R_{e}$}
\def \degsq{deg$^{2}$}
\def \Av{A$_{V}$}
\def \mue{$\bar{\mu}_{e}$}
\def \sigmastar{$\Sigma_{\rm star}$}
\def \msunkpcsq{M$_{\odot}$ kpc$^{-2}$}
\def \lir{L$_{\rm IR}$}
\def \cigale{{\tt CIGALE}}

\begin{document}

\title{Variation of optical and infrared properties of galaxies with their surface brightness}

    \titlerunning{Variation of dust luminosity and attenuation in galaxies with optical surface brightness}

   \author{Junais\inst{1}, 
            K. Ma\l{}ek\inst{1,2},
            S. Boissier\inst{2},
            W. J. Pearson\inst{1},
            A. Pollo\inst{1},
            A. Boselli\inst{2},
            M. Boquien\inst{9},
            D. Donevski\inst{1,10}, 
            T. Goto\inst{3,4},
            M. Hamed\inst{1},             
            S. J. Kim\inst{4},
            J. Koda \inst{5},
            H. Matsuhara\inst{6,7},
            G. Riccio\inst{1},
            M. Romano\inst{1,8}             
            }
    \authorrunning{Junais et al.}
  \institute{National Centre for Nuclear Research, Pasteura 7, PL-02-093 Warsaw, Poland\\ 
             \email{junais@ncbj.gov.pl}
             \and
             Aix Marseille Univ, CNRS, CNES, LAM, Marseille, France   
             \and
             Department of Physics, National Tsing Hua University, 101, Section 2. Kuang-Fu Road, Hsinchu, 30013, Taiwan (R.O.C.) 
             \and
             Institute of Astronomy, National Tsing Hua University, 101, Section 2. Kuang-Fu Road, Hsinchu, 30013, Taiwan (R.O.C.) 
            \and
            Department of Physics and Astronomy, Stony Brook University, Stony Brook, NY 11794-3800, USA 
             \and 
             Department of Space and Astronautical Science, The Graduate University for Advanced Studies, SOKENDAI, 3-1-1 Yoshinodai, Chuo-ku, Sagamihara, Kanagawa 252-5210, Japan 
             \and 
             Institute of Space and Astronautical Science, Japan Aerospace Exploration Agency, 3-1-1 Yoshinodai, Chuo-ku, Sagamihara, Kanagawa 252-5210, Japan 
             \and
             INAF - Osservatorio Astronomico di Padova, Vicolo dell'Osservatorio 5, I-35122, Padova, Italy 
            \and
            Centro de Astronom\'a (CITEVA), Universidad de Antofagasta, Avenida Angamos 601, Antofagasta, Chile 
            \and
            SISSA, Via Bonomea 265, 34136 Trieste, Italy 
             }

   \date{Received 29 March 2023 / Accepted 09 June 2023}

 
  \abstract
  {
  Although it is recognized now that low surface brightness galaxies (LSBs) contribute to a large fraction of the number density of galaxies, many of their properties are still poorly known. LSBs are often considered as ``dust poor'', with a very low amount of dust, based on a few studies.
  
  We use, for the first time, a large sample of LSBs and high surface brightness galaxies (HSBs) with deep observational data to study the variation of stellar and dust properties as a function of the surface brightness/surface mass density.
%
  Our sample consists of 1631 galaxies that are optically selected (with \textit{ugrizy}-bands) at $z<0.1$ from the North Ecliptic Pole (NEP) wide field. We use the large multi-wavelength set of ancillary data in this field, ranging from UV to FIR. We measured the optical size and the surface brightness of the targets, and analyzed their spectral energy distribution using the \cigale{} fitting code. 
  
  Based on the measured average \textit{r}-band surface brightness (\mue{}), our sample consists of 1003 LSBs (\mue{ $>23$} \magperarcsec{}) and 628 HSBs (\mue{ $\leq23$} \magperarcsec{}). We found that the specific star formation rate and specific infrared luminosity (total infrared luminosity per stellar mass) remain mostly flat as a function of surface brightness for both LSBs and HSBs that are star-forming but decline steeply for the quiescent galaxies. The majority of LSBs in our sample have negligible dust attenuation (\Av{ $<0.1$ mag}), except for about 4\% of them that show significant attenuation with a mean \Av{} of 0.8 mag. We found that these LSBs with a significant attenuation also have a high \textit{r}-band mass-to-light ratio ($M/L_r>3$ M$_{\odot}$/L$_{\odot}$), making them outliers from the linear relation of surface brightness and stellar mass surface density.
  These outlier LSBs also show similarity to the extreme giant LSBs from the literature, indicating a possibly higher dust attenuation in giant LSBs as well. 
  
  This work provides a large catalog of LSBs and HSBs with detailed measurements of their several optical and infrared physical properties. 
  Our results suggest that the dust content of LSBs is more varied than previously thought, with some of them having significant attenuation making them fainter than their intrinsic value. With these results, we will be able to make predictions on the dust content of the population of LSBs and how the presence of dust will affect their observations from current/upcoming surveys like JWST and LSST. 
}

   \keywords{Galaxies: general; Galaxies: star formation; Galaxies: ISM}
   \maketitle
%

\section{Introduction}\label{sect:introduction}

In recent years, advances in technology have allowed astronomers to study different types of galaxies in great detail, bringing new interest in low surface brightness galaxies. To have a comprehensive view of galaxy evolution, we have to consider
high surface brightness galaxies (HSBs) and low surface brightness galaxies (LSBs). HSBs are the ``typical'' bright galaxies that have been well-studied in the literature, but LSBs, which are much fainter, have only recently become more accessible for detailed studies. 

LSBs are generally defined as diffuse galaxies that are fainter than the typical night sky surface brightness level of $\sim$23 \magperarcsec{} in the \textit{B}-band \citep{bothun1997}. However, we should note that there is no clear-cut definition for LSBs existing in the literature, and it varies among different works. Therefore, in this work, we consider LSBs as galaxies with an average \textit{r}-band surface brightness \mue{ $>23$} \magperarcsec{} and HSBs with \mue{ $\leq23$} \magperarcsec{}, following similar definitions adopted in previous works \citep[e.g.,][]{martin2019}. 
LSBs span a wide range of sizes, masses, and morphologies, from the most massive giant low surface brightness galaxies (GLSBs) down to the more common dwarf systems \cite[e.g.,][]{Sprayberry1995,Matthews2001,Junais2022}. It is estimated that LSBs make up a significant fraction of more than 50\% of the total number density of the galaxies in the universe \citep{oneil2000,Blanton2005,galaz2011,martin2019}, and about 10\% of the baryonic mass budget \citep{Minchin2004}. Such an abundance of 
LSBs could steepen the faint-end slope of the galaxy stellar mass and luminosity function \citep{sabatini2003,Blanton2005,Sedgwick2019,Kim2022}.
Although LSBs are generally found to be gas-rich, their gas surface densities are usually about a factor 3 lower than for the HSBs \citep{deblok1996,Gerritsen1999}.
As star formation in galaxies is linked to their gas surface density \citep{Kennicutt2012}, this directly affects their ability to form stars, resulting in LSBs having a low stellar mass surface density as well. Therefore, LSBs are a perfect laboratory for studying star formation activity in low-density regimes \citep{boissier2008,wyder2009,bigiel2010}. 

Due to the very low densities and star formation, LSBs are also generally considered to have a very low amount of dust. Their low metallicities also imply that their dust-to-gas ratios should be lower than those of their HSB counterparts \citep{Bell2000}. \citet{Holwerda2005} showed that LSB disks are effectively transparent without any extinction where multiple distant galaxies were observed through their disks. Moreover, most of the observations of LSBs at infrared wavelengths 
resulted in non-detections \citep{hinz2007,rahman2007}, indicating either a very weak or non-detectable dust emission.  

Nevertheless, we cannot necessarily conclude that the entire population of LSBs consisting of a wide range of galaxy types is dust poor. \citet{Liang2010} found that LSBs selected from the SDSS survey span a wide range in their dust attenuation measured using the Balmer decrement (\Av{} in the range of 0 to 1 mag, with a median value of $\sim$0.4 mag). This indicates that not all LSBs are dust poor. However, since surveys like SDSS are very shallow and incomplete beyond \mue{ $>23$} \magperarcsec{}, only the brighter end of the LSB population is observed by them and lack information about the remaining bulk of the faintest LSBs that are missed \citep[e.g.,][]{Kniazev2004,Williams2016}. In another work, \citet{Cortese2012b} showed that the specific dust mass (dust to stellar mass ratio) of local galaxies from the \textit{Herschel} Reference survey (HRS; \citealt{boselli2010}) increases towards fainter galaxies. This yet again indicates that LSBs could have dust masses comparable with HSBs of similar stellar mass. It is likely that the dust in LSBs is distributed very diffusely, similar to their stellar population and gas content, making it extremely hard to detect \citep{Hinz2008}.

%
%

Currently, most studies on dust/infrared properties of LSBs were done using either very small samples \citep[e.g.,][]{rahman2007,hinz2007,wyder2009} or shallow data \citep[e.g.,][]{Liang2010}, which may be not sufficient to make a general conclusion on the large population of LSBs. We need to have a large statistical sample of galaxies at different surface brightness levels to properly understand how these properties change between LSBs and HSBs. In this work, we aim to do this by collecting a large sample of both LSBs and HSBs with deep data to constrain their optical/infrared properties and quantify how the presence of dust (if any) affects our observations of them. Such a work will be particularly significant in the context of current/upcoming observational facilities, such as the Large Synoptic Survey Telescope (LSST; \citealt{ivezic2019}) and the James Webb Space Telescope (JWST; \citealt{Gardner2006}), where a large number of LSBs will be observed.

This paper is structured as follows: Section \ref{sect:data_and_sample} describes the data and the sample used in this work. 
Section \ref{sect:comparison_sample} introduces the comparison sample we use from the literature. Section \ref{section:sed_fitting} describes our spectral energy distribution fitting procedure. The results of our analysis are presented in Sect. \ref{sect:results}, and a global discussion is given in Sect. \ref{sect:discussion}. We conclude in Sect. \ref{sect:conclusion}.

Throughout this work, we adopt a \citet{Chabrier2003} initial mass function (IMF), and a $\Lambda$CDM cosmology with $H_0 = 70 \,\,\text{km s}^{-1} \text{Mpc}^{-1}$,  $\Omega_{M} = 0.27$ and $\Omega_{\Lambda} = 0.73$. All the magnitudes given in this paper are in the AB system.

\section{Data and samples}\label{sect:data_and_sample}

\subsection{Main sample}

In this work, we use the large set of multi-wavelength data ranging from UV to FIR wavelengths available for the North Ecliptic Pole (NEP) wide field, covering an area of $\sim$5.4 \degsq{} (see \citealt{kim2021} for a detailed description of the available data). 
This also includes deep optical data from the Subaru Hyper Suprime-Cam (HSC; \citealt{oi2021}) and CFHT Megcam/Megaprime\footnote{The CFHT Megcam/Megaprime observations of the NEP field covers only a total area of $\sim$3.6 \degsq{}, compared to the $\sim$5.4 \degsq{} covered by the HSC observations.} \citep{huang2020}, which will be used as a basis for our sample selection discussed in Sect. \ref{sample_selection}. The NEP wide field has a very deep coverage in optical with a $5\sigma$ detection limit of 25.4, 28.6, 27.3, 26.7, 26.0, and 25.6 mag in the \textit{ugrizy}-bands, respectively\footnote{Note that at this depth, many local bright galaxies are saturated in the HSC observations and were removed as flagged sources with bad pixels \citep{huang2021}.}. This is very close to the $5\sigma$ depth of the upcoming LSST survey in similar bands \citep{bianco2022}. In both cases, the depth of the data is suited to explore the properties of galaxies as a function of surface brightness, which is the goal of this work.
Moreover, the NEP field is also well suitable for the study of dust and attenuation within galaxies, due to the extensive coverage of this field in the infrared wavelengths (e.g., AKARI, WISE, \textit{Spitzer}, \textit{Herschel}; \citealt{kim2021}) as well as very low foreground Galactic extinction along the line of sight of the NEP field.


\subsubsection{Sample selection}\label{sample_selection}

Our sample selection was done based on the HSC \textit{grizy}-bands and CFHT \textit{u}-band data \citep{huang2020,oi2021}. Only the galaxies with a $5\sigma$ detection in all these six bands were included in our sample. The \textit{u}-band, with its short wavelength, is more sensitive toward dust attenuation. Therefore, the choice of including a \textit{u}-band detection facilitates secure dust attenuation estimates for our sample, which we intend to do in this work. Moreover, a selection in the \textit{ugrizy} also mimics the upcoming LSST-like observations in the same bands, where there will be a vast discovery space for LSBs. 

We also applied an arbitrary selection in redshift, to include only local galaxies with $z<0.1$. We impose this limit since we aim to study the properties of galaxies as a function of surface brightness, and the cosmological dimming would make us lose the LSB galaxies at high-$z$.
For this purpose, we use the photometric redshifts provided by \citet{huang2021}, or the spectroscopic redshifts, whenever available (see \citealt{kim2021} for more details on the available spectroscopic data). The photometric redshifts from \citet{huang2021} were computed with the {\tt Le Phare} code \citep{Arnouts1999,ilbert2006}, using the \textit{ugrizy}-bands. Moreover, the \textit{Spitzer} IRAC 1 (3.6 $\mu$m) and IRAC 2 (4.5 $\mu$m) bands were also included in the photometric redshift estimation, whenever available. The photometric redshifts attain the accuracy of $\sigma_{z_{\rm p}}=0.06$\footnote{The photometric redshift accuracy $\sigma_{z_{\rm p}}$ from \citet{huang2021} is defined as the normalized median absolute deviation, where $\sigma_{z_{\rm p}} = 1.48\times \mathrm{median}\bigg(\frac{|z_{p} - z_{s}|}{1+z_{s}}\bigg)$, with $z_{p}$ and $z_{s}$ being the photometric and spectroscopic redshifts, respectively.} and a catastrophic outlier rate of 8.6\% \citep{huang2021}. With the above selection procedure based on optical detection and the redshift cut, our sample now contains 1950 galaxies. Among them, only 66 galaxies have spectroscopic redshifts.

We verified that a strict selection based on the \textit{ugrizy} bands as discussed above, does not introduce any bias towards bluer/redder galaxies in our sample. To perform this test, we looked at an alternate sample selection, based only on the HSC \textit{grizy}-bands, in the same area as our \textit{u}-band observations. Such a selection increases our sample size by around $\sim$190 galaxies (among them about 90 galaxies are LSBs) as the HSC \textit{grizy} observations are 2 to 3 orders of magnitude deeper than the CFHT \textit{u}-band. However, we found that such a sample has a very similar distribution in their optical colors as our initial \textit{ugrizy} selected sample (mean $g-r$ color of 0.53 mag for both the samples). This indicates that the inclusion of the \textit{u}-band does not introduce a bias in our selection. Therefore, from hereupon, we chose to continue with our initial \textit{ugrizy} and redshift selected sample of 1950 galaxies.

\subsubsection{Morphological fitting}\label{sect:morphological-fitting}

In order to obtain the effective surface brightness and radius of each galaxy, we performed a morphological fitting procedure using the {\tt AutoProf} tool \citep{stone2021}. {\tt AutoProf} is an efficient tool to capture the full radial surface brightness light profile of a galaxy from its image using a non-parametric approach, unlike the parametric fitting tools like Galfit \citep{peng2002}, which do not always capture the total light from a galaxy. {\tt AutoProf} is also well-suited for low surface brightness science, where it can extract about two orders of magnitude fainter isophotes from an image than any other conventional tool \citep{stone2021}.

\begin{figure}[ht]
\includegraphics[scale=0.65]{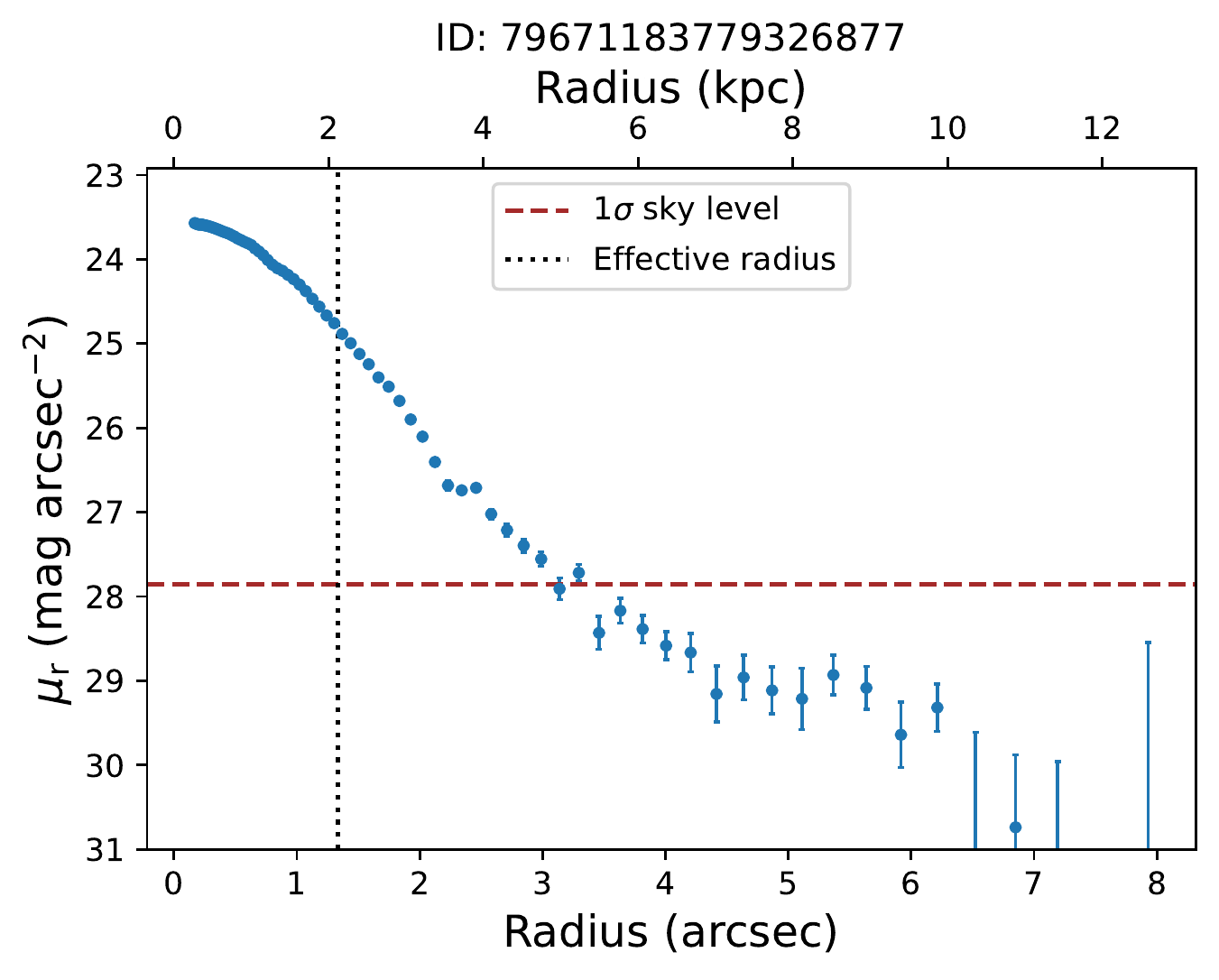}
\centering
\llap{\shortstack{%
        \includegraphics[scale=0.10]{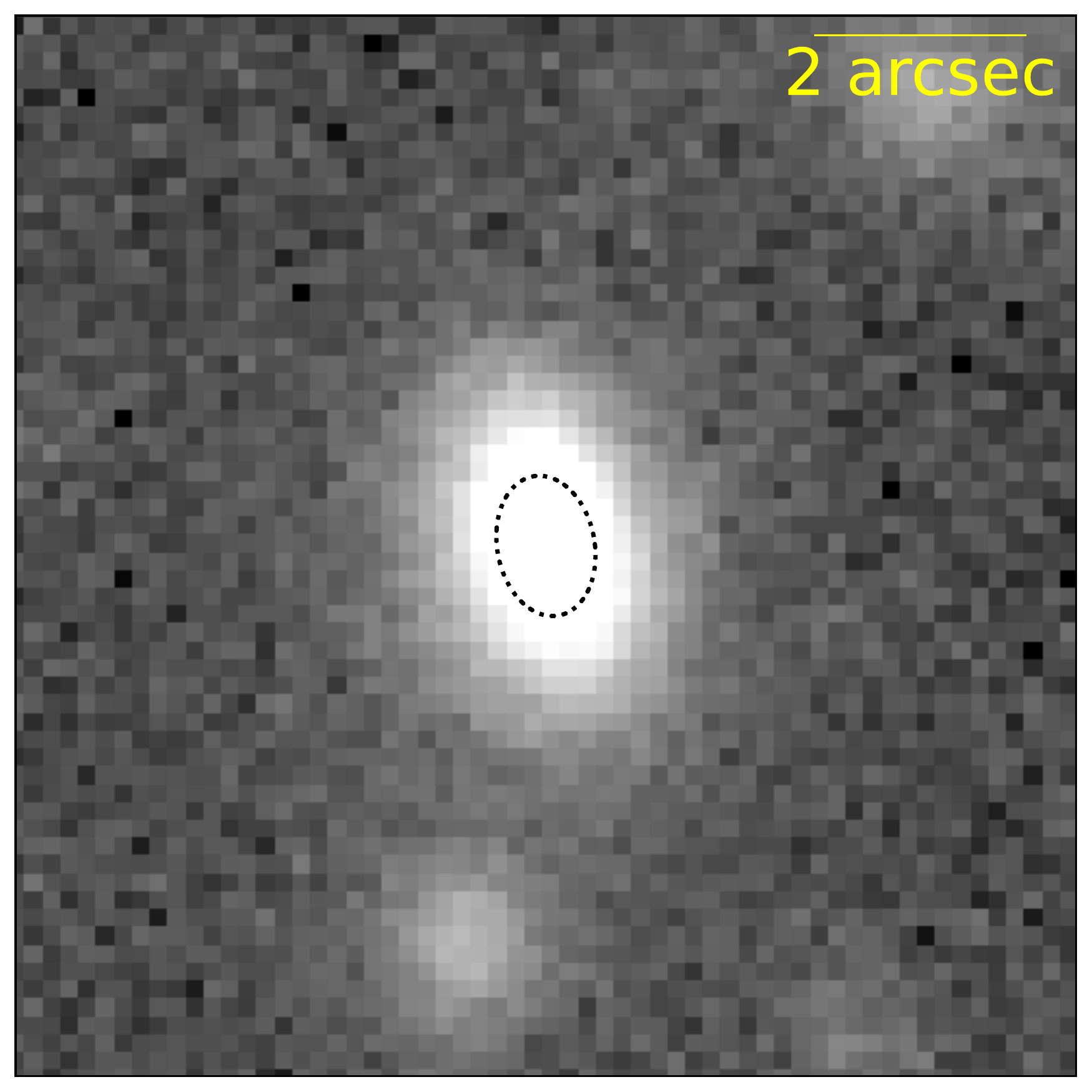}\\
        \rule{0ex}{3.5cm}%
      }
  \rule{0.5cm}{0ex}}
\caption{Example of an \textit{r}-band radial surface brightness profile extracted for a galaxy using {\tt AutoProf}. The HSC \textit{r}-band image of the galaxy is shown as the inset. The black dotted line marks the effective radius of the galaxy. The brown dashed horizontal line is the $1\sigma$ sky noise level.}\label{fig:sb-profile}
\end{figure}
\begin{figure}
    \centering
    \includegraphics[width=\hsize]{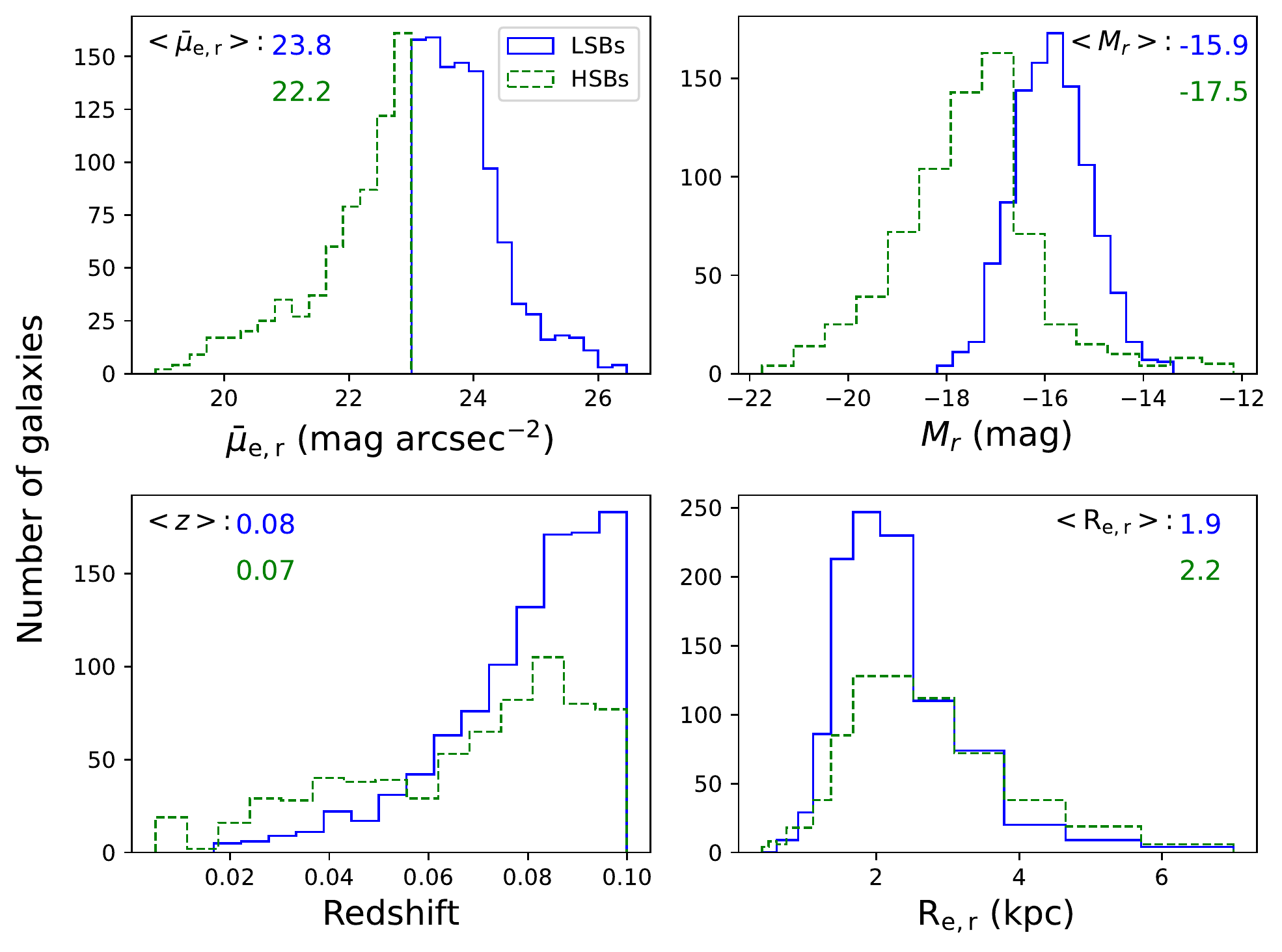}
    \caption{Distribution of the basic properties of the LSBs (blue solid line) and HSBs (green dashed line) in our sample. The average \textit{r}-band surface brightness within the effective radius and the \textit{r}-band absolute magnitude are given in the top left and top right panels, respectively. The
    redshift and the effective radius are in the bottom left and bottom right panels, respectively. The median values corresponding to each parameter are marked inside each panel in blue and green for the LSBs and HSBs, respectively.}
    \label{fig:hist_basic_params}
\end{figure}

The surface brightness profile extraction of our sample was done on the HSC \textit{r}-band images. Although the \textit{g}-band is the deepest among our sample, the choice of the \textit{r}-band (which is the second deepest) is motivated by the fact that \textit{r}-band is a better tracer of the stellar mass distribution in galaxies than the \textit{g}-band \citep{Mahajan2018}.
Figure \ref{fig:sb-profile} shows an example of the surface brightness profile obtained for a galaxy.
Similarly, we extracted the profiles for the majority of the galaxies in our sample (1743 out of 1950 galaxies). The remaining sources have failed/flagged profile fits. Therefore, hereafter we exclude from our sample all the sources without a reliable morphological fit, which leaves 
1743 galaxies. We integrated each surface brightness profile until its last measured radius to estimate the total light from each galaxy, the corresponding effective radius (half-light radius; \Re{}), and the average surface brightness within the effective radius ($\bar{\mu}_{e}$). From Fig. \ref{fig:sb-profile}, we can clearly see that the radial surface brightness profiles we obtained using {\tt AutoProf} reach well beyond the effective radius of the galaxy to about 4 times \Re{} and also $\sim$2 \magperarcsec{} deeper than the typical sky level (a similar trend is found for our full sample), which is ideal to probe low surface brightness galaxies. The distribution of the \textit{r}-band \Re{} and \mue{} for our full sample is given in Fig. \ref{fig:hist_basic_params}. Our sample at this stage consists of 1041 LSBs and 702 HSBs (although such a distinction is based on an arbitrary definition as discussed in Sect. \ref{sect:introduction}). The LSBs have a median \mue{} and \Re{} of 23.8 \magperarcsec{} and 1.9 kpc, respectively. Whereas the HSBs are brighter and slightly larger in size with a median \mue{} and \Re{} of 22.2 \magperarcsec{} and 2.2 kpc, respectively. In terms of the redshift, both the LSBs and HSBs have a similar distribution, with a median value of about 0.08. The \textit{r}-band absolute magnitudes ($M_{r}$) of the two sub-samples show a clear difference, with the LSBs fainter than the HSBs, as expected from their selection, with a median $M_{r}$ of $-15.9$ mag and $-17.9$ mag, respectively. 


We also made a comparison of our \Re{} and $\bar{\mu}_{e}$ estimates with that of \citet{Pearson2022}, who made a S\'ersic profile fitting of the NEP galaxies in the same band using the {\tt statmorph} tool \citep{Rodriguez-Gomez2019}. We found that, in general, our values are in agreement with \citet{Pearson2022}, with a mean difference in \Re{} of $0.01\pm0.23$ dex and $-0.02\pm0.76$ \magperarcsec{} for the \mue{}.  


\subsubsection{Cross-matching with multi-wavelength catalogs}\label{sect:crossmatch-catalogs}

\begin{figure}
    \centering
    \includegraphics[width=\hsize]{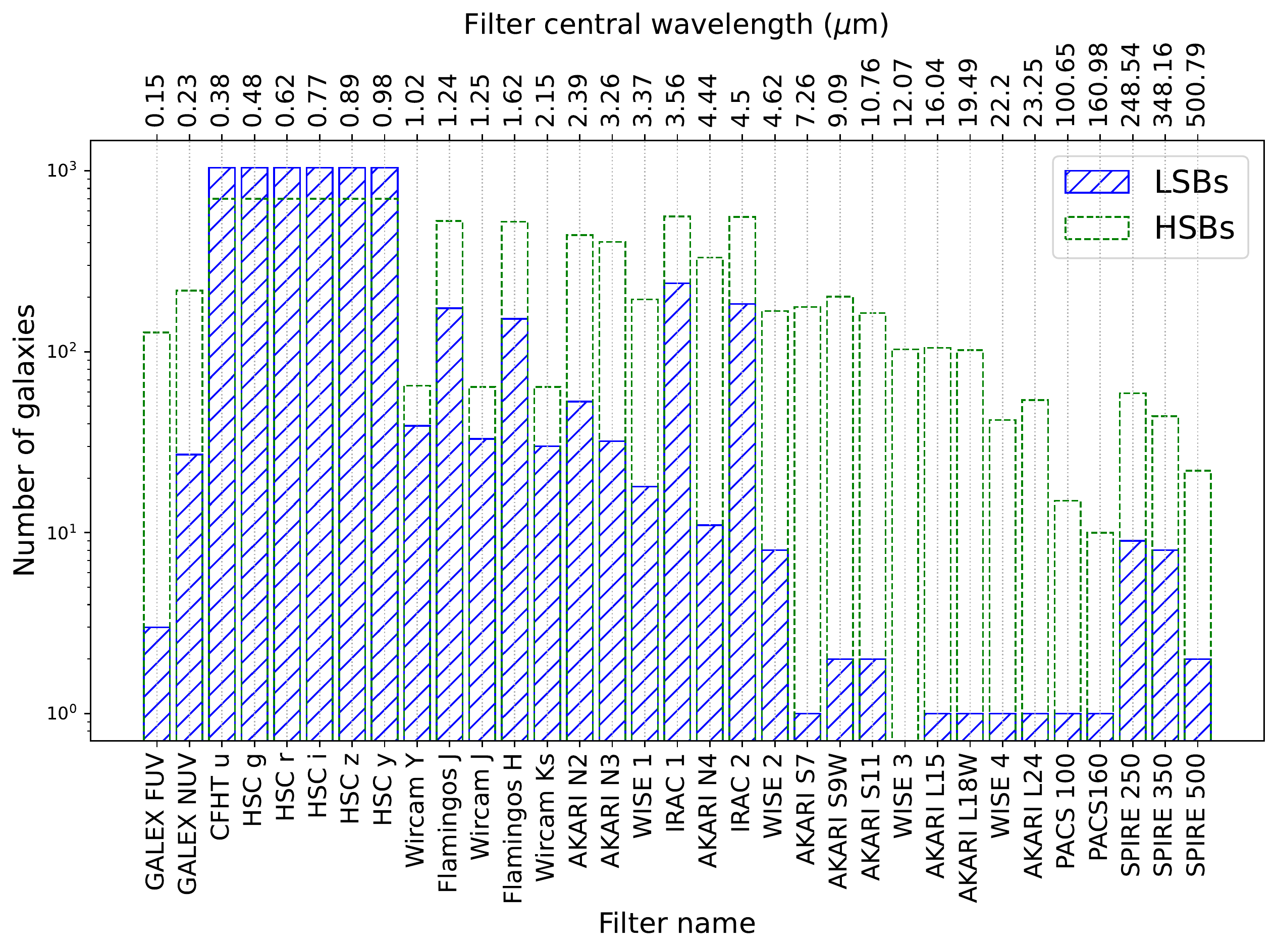}
    \caption{Distribution of the multi-wavelength data available for the sample. The LSBs and HSBs are marked as the blue striped bars and the green dashed bars, respectively. The broadband filter names and their corresponding wavelengths are given in the bottom and top horizontal axes, respectively. By selection, all the galaxies in the sample have detection in the \textit{ugrizy}-bands.}
    \label{fig:multiband-data-histogram}
\end{figure}

After the initial sample selection and their morphological fitting, we cross-matched our optically selected sample with all the available multi-wavelength data in hand. For the NEP field, other than the optical data from HSC and CFHT, we have ancillary data available from GALEX (FUV and NUV bands; \citealt{Bianchi2017}), AKARI (N2, N3, N4, S7, S9W, S11, L15, L18W, and L24 bands; \citealt{kim2012}), CFHT/WIRCam (\textit{Y}, \textit{J}, and $K_s$ bands; \citealt{Oi2014}), KPNO/FLAMINGOS (\textit{J} and \textit{H} bands; \citealt{jeon2014}), \textit{Spitzer}/IRAC (band 1 and 2; \citealt{nayyeri2018}), WISE (band 1 to 4; \citealt{jarrett2011}) and \textit{Herschel} PACS/SPIRE (100 $\mu$m, 160 $\mu$m, 250 $\mu$m, 350 $\mu$m and 500 $\mu$m bands; \citealt{pearson2017,pearson2019}). A detailed description of the data is given in \citet{kim2021}. The multi-band photometry obtained from the cross-matching of these catalogs will be used in the spectral energy distribution (SED) fitting procedure discussed in Sect. \ref{section:sed_fitting}. The cross-matching was done following \citet{kim2021}, where a $3\sigma$ positional offsets in the RA/Dec. coordinates corresponding to each dataset with respect to the HSC coordinates were used as the cross-matching radii. For GALEX, AKARI, WIRCam, FLAMINGOS, IRAC, WISE, PACS, and SPIRE, we used a cross-matching radius of 1.5\arcsec, 1.5\arcsec, 0.5\arcsec, 0.65\arcsec, 0.58\arcsec, 0.7\arcsec, 2.75\arcsec, and 8.44\arcsec, respectively. Figure \ref{fig:multiband-data-histogram} shows the distribution of galaxies with counterparts in each dataset. About 62\% of galaxies in the sample (1086 out of 1743 sources) have at least one counterpart outside the \textit{ugrizy} optical range.

We also compared our sample with the band-merged catalog of \citet{kim2021} who identified HSC counterparts for the AKARI-detected sources in NEP. Only 532 galaxies of our sample overlap with the \citet{kim2021} catalog, indicating that the remaining of our sources do not have any AKARI counterparts in NIR or MIR. Moreover, $\sim$85\% of our sample does not have any detection in the mid-infrared (MIR) and far-infrared (FIR) regime (in the 7 $\mu$m to 500 $\mu$m wavelength range) as shown in Fig. \ref{fig:multiband-data-histogram}. However, since we aim to study the IR properties of our sample, it is crucial to have observational constraints in the MIR and FIR range. We have deep observations from AKARI and \textit{Herschel}/SPIRE in this wavelength range, covering the entire field we study. Therefore, for the galaxies without any detection in this range, 
we use the detection limits from these observations as their flux upper limits\footnote{We used the AKARI and \textit{Herschel}/SPIRE upper limits given in Table 1 of \citealt{kim2021}, as they have the deepest coverage in the entire NEP Wide field for the MIR and FIR range.}. The $5\sigma$ detection limits of the AKARI S7, S9W, S11, L15, L18W, L24, and SPIRE 250 $\mu$m, 350 $\mu$m and 500 $\mu$m bands, are 0.058 mJy, 0.067 mJy, 0.094 mJy, 0.13 mJy, 0.12 mJy, 0.27 mJy, 9 mJy, 7.5 mJy, and 10.8 mJy, respectively \citep{kim2021}. These upper limits are used in the SED fitting procedure discussed in Sect. \ref{section:sed_fitting}.

\subsection{Comparison sample}\label{sect:comparison_sample}

We use the \textit{Herschel} Reference Survey (HRS; \citealt{boselli2010}) sample for the comparison of the results obtained in this work. HRS is a volume-limited sample ($15\leq D \leq 25$ Mpc) of 322 galaxies consisting of both early-type and late-type galaxies (62 early-type galaxies with \textit{K}-band magnitude $K_s\leq8.7$ mag and 260 late-type galaxies with $K_s\leq12$ mag). The HRS sample is selected in such a way as to include only the high galactic latitude ($b>+55\degree$) sources with low Galactic extinction (similar to the NEP sample). The HRS sample covers a large range of galaxy properties and therefore it can be considered a representative sample of the local universe. A detailed description of the HRS sample is provided in \citet{boselli2010}.

We make use of the extensive studies done in the literature on this sample \citep[e.g.,][]{Cortese2012a,Cortese2012b,Ciesla2014,Boselli2015,Andreani2018} for comparison purposes. The optical structural properties (\textit{r}-band \Re{} and $\bar{\mu}_{e}$) and the stellar masses of the HRS sample used in this work are taken from \citet{Cortese2012a,Cortese2012b}. The star formation rates (SFR) and the \textit{V}-band dust attenuation values (\Av{}) are provided by \citet{Boselli2015}, with the SFR estimated as the combined average of multiple star formation tracers ranging from UV to FIR and radio continuum. Only about 200 late-type galaxies in the HRS sample have available \Av{} measurements which we use in this work. The \Av{} of the HRS galaxies are computed from the Balmer decrement.
The total infrared luminosity (\lir{}) for all the HRS sources is taken from \citet{Ciesla2014}, who used SED fitting method to estimate the \lir{}, similar to the approach we use in this work. Since the HRS also includes galaxies in the Virgo cluster, where dust can be stripped away during the interaction of galaxies with their surrounding environment, the dust content of such galaxies is principally regulated by external effects rather than secular evolution. Therefore, we removed from our comparison the HRS galaxies with a large HI gas deficiency parameter ($HI-def > 0.4$), which is an indicator of environmental interactions \citep{Boselli2022}. Our HRS comparison sample now consists of 159 galaxies. A detailed description of the compilation of the HRS data is given in \citealt{Andreani2018}. 

Although the HRS is a \textit{K}-band selected sample, it is a well-studied local sample of galaxies with high-quality data. Therefore, throughout this work, we use the HRS as a control sample from the literature to compare and validate our results.





\section{SED fitting}\label{section:sed_fitting}

\subsection{Method}

We used the Code Investigating GAlaxy Emission (\cigale{}\footnote{\url{https://cigale.lam.fr/2022/07/04/version-2022-1/}}; \citealt{noll2009,Boquien2019}) SED fitting tool to estimate the physical parameters of the galaxies in our sample, in particular, the stellar mass, SFR, total infrared luminosity and dust attenuation. \cigale{} uses an energy balance principle where the stellar emission in a galaxy is absorbed and re-emitted in the infrared by the dust. This enables us to simultaneously fit the UV to FIR emission of the galaxies in our sample. The input parameters we used for our SED fitting are given in Table \ref{table:cigale_params}. 

We use the \citet{bruzual-charlot2003} stellar population synthesis models with a \citet{Chabrier2003} IMF and a fixed sub-solar stellar metallicity of 0.008 (0.4 $Z_{\odot}$)\footnote{Adopting different metallicity values were found to have only a negligible impact on the overall results presented in this paper. Therefore, as we focus mainly on LSBs, we chose to keep the metallicity at a sub-solar value to reduce the number of free parameters in our fitting procedure.}. 
We also adopted a flexible star formation history (SFH) from \citet{Ciesla2017} which includes a combination of delayed SFH with the possibility of an instantaneous recent burst/quench episode. Such an SFH was successfully used to reproduce a broad range of galaxy properties in the local universe \citep{Hunt2019,Ciesla2021}. The range of values adopted for the SFH is given in Table \ref{table:cigale_params}.

We also include dust attenuation, adopting 
the \texttt{dustatt\_modified\_starburst} module of \cigale{}, which is a modified version of the well-known \citet{Calzetti2000} attenuation curve, extended with the \citet{Leitherer2002} curve between the Lyman break and 150 nm. This module also provides a possibility of changing the slope as well as the addition of a UV bump in the attenuation curve. In this work, we 
fix these parameters to their standard value to reduce the number of free parameters as we have only 6 photometric bands for a large fraction of our sample.
The \texttt{dustatt\_modified\_starburst} module treats the stellar continuum and the emission lines differently, with the latter being attenuated more by the dust (this difference in attenuation of the continuum and the lines is controlled by a factor, which is kept as a constant, as shown in Table \ref{table:cigale_params}). The color excess of the lines, $E(B-V)_{\rm lines}$
, is left as a free parameter with values ranging from 0 to 2 mag. 

Once the dust attenuation is modeled, we need to use a dust emission module to model the re-emission of the attenuated radiation in the MIR to FIR. For this purpose, we adopted the \citet{Dale2014} dust emission models based on nearby star-forming galaxies. The \citet{Dale2014} models only have two free parameters (AGN fraction and the slope of the radiation field intensity, $\alpha$). Since only less than 0.5\% of local dwarf galaxies possess an AGN \citep{Reines2013,Lupi2020}, in this work, we assume an AGN fraction of zero for our sample as it mostly consists of low-mass galaxies with a median \textit{r}-band absolute magnitude of the order of $-17$ mag, as shown in Fig. \ref{fig:hist_basic_params}. 
%
For the slope of the radiation field intensity, we use a fixed value\footnote{We verified that a variation of the interstellar radiation field slope $\alpha$ from 2 to 3 does not make any significant change (a change of less than 0.1 dex on all our estimated quantities) in our SED results.} of $\alpha=2$. 

\begin{figure*}
    \centering
    \includegraphics[width=\hsize]{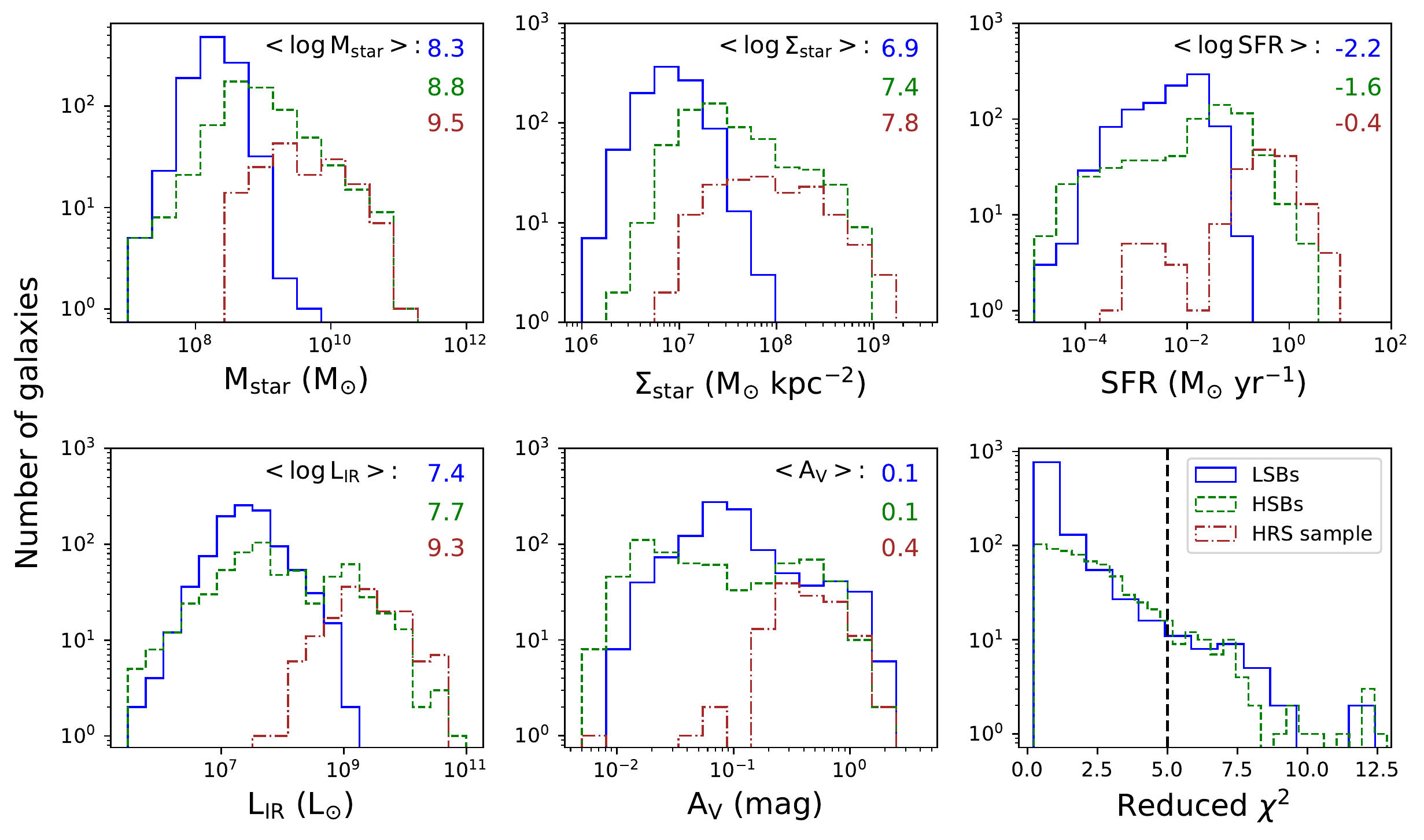}
    \caption{Distribution of the best-fit parameters of our NEP sample obtained from the SED fitting. The LSBs and HSBs are marked as the blue solid and the green dashed histograms, respectively. The HRS comparison sample used in this work is shown as the brown dash-dotted distribution. The median values corresponding to each parameter are marked inside the panels in blue, green and brown colors for the LSBs, HSBs and the HRS sample, respectively. The bottom right panel gives the reduced \chisq{} obtained from the SED fitting, along with the black vertical dashed line marking the arbitrary selection cut we used to remove bad fits.}
    \label{fig:hist_sed_best_params}
\end{figure*}

We performed the SED fitting of our sample with over 130 million models ($\sim$\numprint{200000} models per redshift). For the galaxies without any detection in the MIR/FIR regime (>7 $\mu$m), we use the $5\sigma$ flux upper limits discussed in Sect. \ref{sect:crossmatch-catalogs}. These upper limits are important in constraining the IR properties of our optically selected galaxies. \cigale{} treats the upper limits
in a mathematically correct way to compute the total \chisq{} of a SED.
After the SED fitting, we obtained a median reduced \chisq{} value ($\chi^2_{r}$) of 0.95 with an absolute deviation of 0.64 (see Fig. \ref{fig:hist_sed_best_params}). About 94\% of the sample (1631 out of 1743 galaxies) has $\chi^2_{r}$ less than an arbitrary value of 5. From hereupon, we exclude all the remaining sources with $\chi^2_{r} > 5$ for our further analysis. Our final sample now consists of 1631 galaxies (1003 LSBs and 628 HSBs).

\begin{figure*}
    \centering
    \includegraphics[width=0.49\hsize]{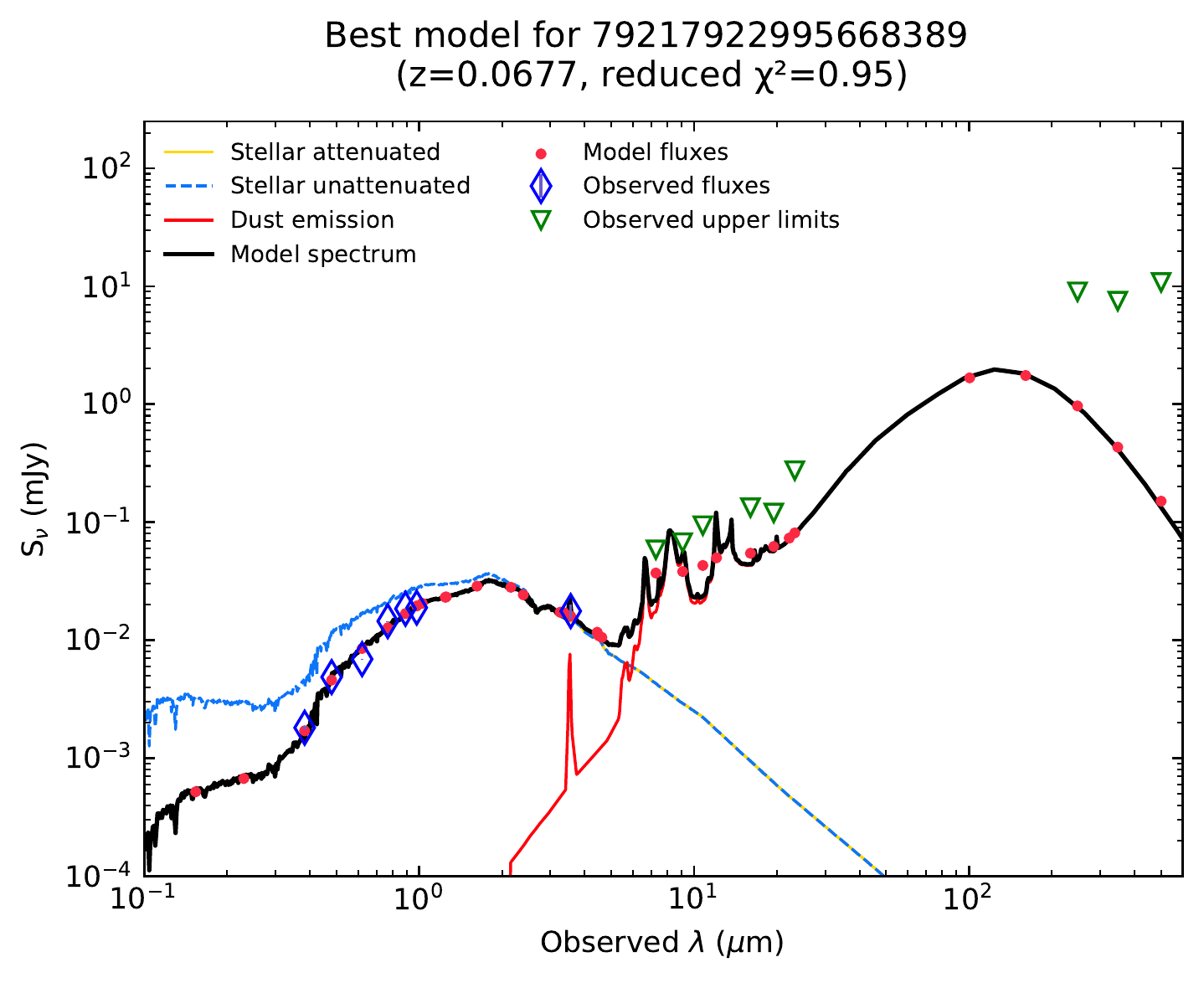}
    \includegraphics[width=0.49\hsize]{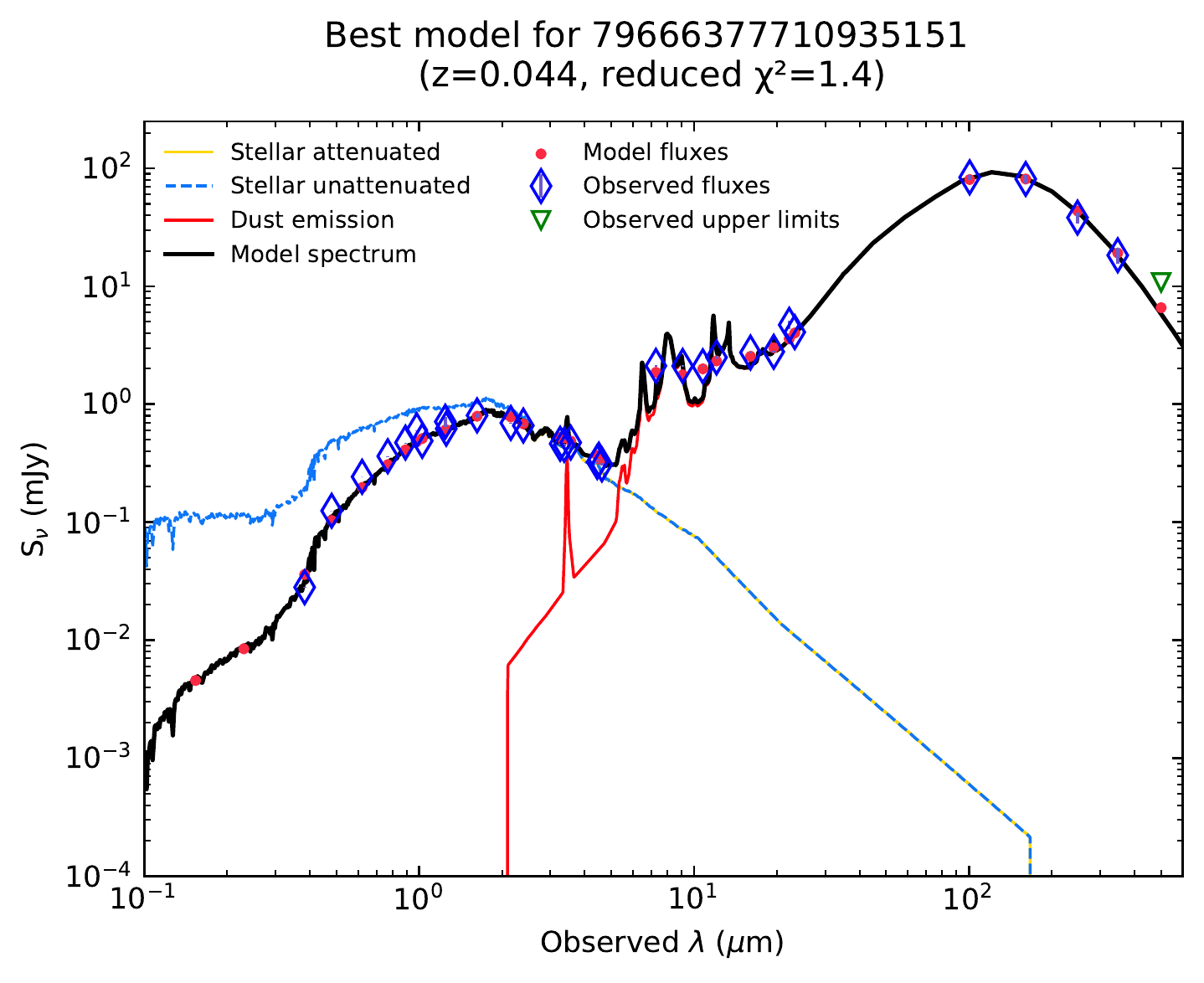}
    \caption{Examples of our best-fit SEDs. The left panel shows the best-fit SED for a low surface brightness galaxy with only upper limits in the MIR and FIR wavelengths, whereas the SED in the right panel is of a high surface brightness galaxy with extensive photometry at all wavelengths. The black solid line is the model SED. The blue open diamonds and the red circles are the observed fluxes and best-fit model fluxes, respectively. The green downward triangles are the observed flux upper limits used in the SED fitting.}
    \label{fig:sed_bestmodel}
\end{figure*}

Fig. \ref{fig:sed_bestmodel} gives an example of the best-fit SEDs obtained for an LSB and HSB galaxy. We can see that for both galaxies we obtain a good fit, with the upper limits providing strong constraints on the IR emission of the galaxy without any MIR/FIR detection. 

\subsection{Robustness of the SED fitting results}\label{sect:robustness}

The robustness of the estimated physical parameters from our SED fit was verified by several tests. For each parameter, throughout this work, we used the Bayesian mean and standard deviation of the quantities estimated by \cigale{}, based on the probability distribution function of the tested models. This ensures a more robust estimate of a quantity and its uncertainty, rather than directly using the best-fit model parameter, especially in case of degeneracy between physical parameters. 

Another feature we used to check the reliability of the estimated parameters is by using a mock analysis provided by \cigale{}. In this test, \cigale{} builds a mock catalog with synthetic fluxes for each object based on its best-fit SED. The synthetic fluxes of each filter are modified by adding a random noise based on the uncertainty of the observed fluxes in the corresponding filter. Later, \cigale{} performs the same calculations on this mock catalog as done for the original observations to get the mock physical parameters. The results of the mock analysis are given in Appendix \ref{appendix:sect_mock_analysis}. We see that the stellar mass is the most well-constrained quantity, with the square of the Pearson correlation coefficient ($r^2$) equals 0.99, followed by the total infrared luminosity ($r^2=0.85$), the SFR ($r^2=0.81$) and the \textit{V}-band attenuation ($r^2=0.77$).
Although the SFR, \lir{} and \Av{} have a larger scatter (0.41 dex, 0.31 dex, and 0.14 mag, respectively),
based on the linear regression analysis shown in Fig. \ref{appendix:fig_mocks}, we can still consider them reliable as estimates. 

We also performed yet another test to verify the robustness of our estimated physical quantities. A separate SED fitting, similar to our original fits, was done for only the FIR-detected galaxies in our sample (53 galaxies with detection in either \textit{Herschel} PACS/SPIRE), but this time only using their optical \textit{ugrizy}-bands photometry. This was done to check how well we can recover the ``true'' quantities by only using the \textit{ugrizy} photometry. We compared the results of this fit with our original fit results and found that for the FIR-detected galaxies, the \mstar{}, SFR, \lir{} and \Av{} obtained from the original fit and the optical-only fit have a mean difference of $-0.09$ dex, $-0.26$ dex, $-0.24$ dex and $-0.07$ mag, respectively, as given in Fig. \ref{appendix:fig_fir_ugrizy}. The negative values indicate that a fit using only optical bands (or galaxies with only optical detection) in general has overestimated quantities but only by a few tenths of dex. We verified that this trend remains the same for our entire sample if we perform the SED fitting without using any flux upper limits in the MIR and FIR. Similarly, we examined how a change in our upper limit definitions from 5$\sigma$ to 2$\sigma$ in the SED fitting affects our results. Such a change only has a negligible effect on our overall results with the stellar mass being unchanged and the SFR, \lir{} and \Av{} changed by only 0.05 dex, 0.16 dex and 0.02 mag.

Table \ref{appendix:table_estimated_properties} provides all the estimated parameters of our sample.   





\begin{table*}
        \centering
	\caption{Input parameters for \cigale{} SED fitting}
	\label{table:cigale_params}
	\begin{tabular}{ll}
		\hline
		Model and Input parameters & Values \\ 
		\hline
		\hline\\
		\textbf{Star-formation history}: \texttt{sfhdelayedbq} \citep{Ciesla2017}\\ 
		\hspace{4pt} e-folding time of the main stellar population model (Myr) & 500, [1000,10000] with a spacing of 1000 \\
		\hspace{4pt} Age of the main stellar population in the galaxy (Myr) & [10000,13000] with a spacing of 500 \\
		\hspace{4pt} Age of the burst/quench episode (Myr) & 100, 200, 400, 600, 800, 1000 \\
		\hspace{4pt} Ratio of the SFR after and before the burst/quench (Myr) & 0,0.2,0.4,0.6,0.8,1,1.2,1.4,1.6,1.8,2 \\
            \hline\\
		\textbf{Stellar population}: \texttt{bc03} \citep{bruzual-charlot2003}\\ 
		\hspace{4pt} Initial mass function & \citet{Chabrier2003} \\ 
		\hspace{4pt} Metallicity & 0.008 \\
		\hline\\
  
		\textbf{Dust attenuation}: \texttt{dustatt\_modified\_starburst} \\ 
            \citep{Calzetti2000,Leitherer2002}\\		
		\hspace{4pt} $E(B-V)_{\rm lines}$, the color excess of the nebular lines (mag)& 0, [0.001,2] log sampled with 40 values\\
		\hspace{4pt} Reduction factor to compute $E(B-V)_{\rm continuum}$  & 0.44 \\
		\hspace{4pt} Amplitude of the UV bump & 0.0 \\
            \hspace{4pt} Slope delta of the power law modifying the attenuation curve & 0.0 \\
		\hspace{4pt} Extinction law for attenuating emission lines flux & Milky Way \citep{Cardelli1989}\\
		\hspace{4pt} R$_{\rm V}$ & 3.1 \\
  
		\hline\\
		\textbf{Dust emission}: \texttt{dale2014} \citep{Dale2014}\\ 
		\hspace{4pt} AGN fraction & 0.0 \\
		\hspace{4pt} Slope of the interstellar radiation field ($\alpha$) & 2.0 \\
		\hline
	\end{tabular}
\end{table*}

\section{Results}\label{sect:results}

Figure \ref{fig:hist_sed_best_params} shows the distribution of several physical parameters (stellar mass, stellar mass surface density, SFR, \lir{} and \Av{}) obtained after the SED fitting discussed in Sect. \ref{section:sed_fitting}. Our sample predominantly consists of low-mass galaxies with both the LSBs and HSBs having a median stellar mass of $10^{8.3}$ \msun{} and $10^{8.8}$ \msun{}, respectively. The HRS sample lies along the massive end of the distribution with a median stellar mass of $10^{9.5}$ \msun{}. Using the stellar mass and the measured radius (as discussed in Sect. \ref{sect:morphological-fitting}), we estimated the stellar mass surface densities (\sigmastar{}) of our sample following \citet{Cortese2012b} as shown in Eq. \ref{eqn:sigma_star}:

\begin{equation}\label{eqn:sigma_star}
    \Sigma_{\rm star} = \frac{M_{\rm star}}{2\pi R_{e}^{2}}, 
\end{equation}
where $R_{e}$ is the \textit{r}-band half-light radius and \mstar{} is the stellar mass. 
Equation \ref{eqn:sigma_star} is a widely used method in the literature to estimate \sigmastar{} for both LSBs and HSBs \citep[e.g.,][]{Kauffmann2003,Zhong2010,Cortese2012b,Grootes2013,Boselli2022,Carleton2023}. Although several other methods also exist to obtain \sigmastar{}, many of them provide similar values without changing the global properties of our sample. For instance, we tried estimating \sigmastar{} following \citet{Chamba2022}\footnote{Following \citet{Chamba2022}, \sigmastar{} of our sample can also be estimated as:$$\log \Sigma_{\rm star} \,(\rm M_{\odot}\,\rm pc^{-2}) = 0.4\times(M_{r,\odot} - \mu_{r}) + \log M/L_{r} + 8.629,$$ where $M_{r,\odot}$ is the absolute magnitude of the sun in the \textit{r}-band filter ($M_{r,\odot} = 4.64$ mag for HSC \textit{r}-band), $\mu_{r}$ and $M/L_{r}$ are the \textit{r}-band surface brightness and stellar mass-to-light ratio, respectively.} using our observed \mue{} and the stellar mass-to-light ratio (M/L) obtained from the SED fitting (ratio of the stellar mass and the observed \textit{r}-band luminosity). This method does not rely on the measured \Re{} values as in Eq. \ref{eqn:sigma_star}. We found that the \sigmastar{} estimates from both methods are similar with a mean difference of $-0.15$ dex (in general, the second method gives a slightly higher \sigmastar{}). However, we should note that the above two methods only provide an average value of the \sigmastar{} of a galaxy, and therefore such minor differences connected to the adopted methodology can be neglected. Estimating the ``true'' value of \sigmastar{} requires resolving individual stellar populations as well as information on the radial distribution of dust that can affect \sigmastar{} measurements. With our current data in hand, it is beyond the scope of this work. Therefore, from hereupon, we adopt the \sigmastar{} values estimated using the simple and widely used method from Eq. \ref{eqn:sigma_star}, and their distribution is shown in Fig. \ref{fig:hist_sed_best_params}.

The \sigmastar{} also follow a similar distribution as the stellar mass with the LSBs and HSBs having a median \sigmastar{} of $10^{6.9}$ \msunkpcsq{} and $10^{7.4}$ \msunkpcsq{}, respectively, whereas the HRS sample with a value of $10^{7.8}$ \msunkpcsq{}. In terms of the star formation rate, LSBs and HSBs have a median SFR of $10^{-2.2}$ \msun{} yr$^{-1}$ and $10^{-1.6}$ \msun{} yr$^{-1}$, respectively, and the HRS galaxies with a corresponding value of $10^{-0.4}$ \msun{} yr$^{-1}$. The \lir{} shows a similar distribution as the SFR, with the LSBs, HSBs, and the HRS galaxies having median values of $10^{7.4}$ \lsun{}, $10^{7.7}$ \lsun{} and $10^{9.3}$ \lsun{}, respectively. Figure \ref{fig:hist_sed_best_params} also shows the distribution of the \Av{}. For both the LSBs and HSBs, we find a median \Av{} of 0.1 mag, with \Av{} values ranging from almost zero to 2 mag. The HRS sample has a higher median \Av{} of 0.4 mag.
From the above comparison of the physical parameters, we can see that, our sample extends towards the low-mass regime, as well as lower SFR, \lir{} and \Av{}, much more than the HRS sample. 

In the following subsections, we investigate the dependence of these quantities as a continuous function of \sigmastar{}, in an attempt to understand how the geometrical distribution of stars within galaxies affects their global parameters. We choose the \sigmastar{} over \mue{} for our comparisons due to several reasons. The \sigmastar{} is a widely used quantity in the literature to compare galaxy physical properties and provides a more intrinsic physical quantity than \mue{}. Moreover, although \mue{} is a directly observed quantity, its value depends highly on the choice of an observed filter, whereas \sigmastar{} is less affected by that. In the Sect. \ref{sect:mue_sigmastar} we show a comparison of the \mue{} and \sigmastar{} our sample.




\subsection{Optical surface brightness}\label{sect:mue_sigmastar}

\begin{figure}
    \centering
    \includegraphics[width=\hsize]{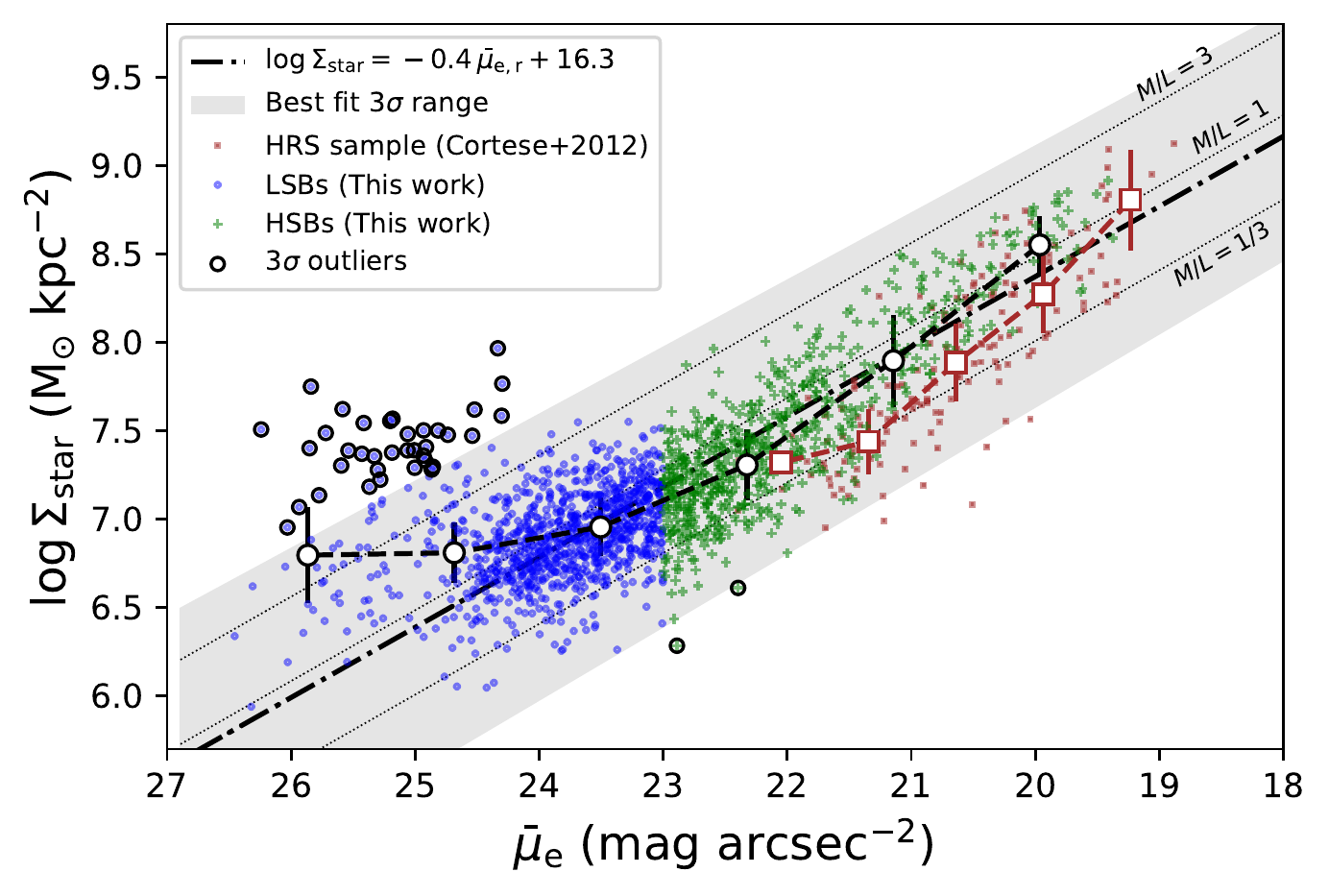}
    \caption{Stellar mass surface density (\sigmastar{}) as a function of the \textit{r}-band average surface brightness within the effective radius (\mue{}). The LSBs and HSBs are marked as blue open circles and green crosses, respectively. The black dashed line and the black circles mark the median distribution of our sample (the error bars are the median absolute deviation). The \mue{} and \sigmastar{} has a mean uncertainty of are 0.03 \magperarcsec{} and 0.07 \msunkpcsq{}, respectively. The brown squares are the HRS sample from \citet{Cortese2012a}, and the brown dashed line and squares are their median distribution. The black dash-dotted line and the grey shaded region are the linear best-fit and $3\sigma$ scatter of our sample, respectively. The black open circles around some sources are the 3$\sigma$ outliers of the best-fit line. The three black dotted lines correspond to the expected relation between \sigmastar{} and \mue{} based on fixed fiducial mass-to-light ratios of 1/3, 1 and 3 M$_{\odot}$/L$_{\odot}$ \citep{Chamba2022}, as discussed in Sect. \ref{sect:results}.}
    \label{fig:mue_sigmastar}
\end{figure}

The surface brightness of a galaxy is the distribution of its stellar light per unit area. It is related to the total stellar mass surface density of a galaxy in the same way as galaxy luminosity and stellar mass are related by their mass-to-light ratio. Although there are several relations in the literature that explores the connections between galaxy surface brightness, luminosity, and stellar mass \citep[e.g.,][]{Boselli2008, martin2019, Jackson2021}, there exists a large scatter among such relations. For instance, \citet{Jackson2021} illustrates that for a fixed stellar mass, galaxies show a large scatter in their surface brightness up to $\sim$3 \magperarcsec{}, ranging from LSBs to HSBs. Although it is well known that the stellar mass is one of the main drivers of galaxies' properties \citep[e.g.,][]{Kauffmann2003,Speagle2014}, considering that a large scatter exists at any given stellar mass for the surface brightness, it is important to explore the possible trends in surface brightness associated with other quantities. In Fig. \ref{fig:mue_sigmastar} we explore such a relation using our observed \mue{} and the stellar mass surface density (\sigmastar{}).

Our sample covers a large range of surface brightness ($\sim$7 order of magnitudes) and stellar mass surface densities (3 dex) from bright to very faint galaxies. This is about 4 orders of magnitude deeper in surface brightness than the HRS sample. For the HSBs (\mue{ $<23$} \magperarcsec{}), the \sigmastar{} follows a linear trend with \mue{}, consistent with the observations from the HRS sample. However, for the LSBs (\mue{ $>23$} \magperarcsec{}, which the HRS sample does not probe), the brighter tail (23 $<$ \mue{} $< 24.5$ \magperarcsec{}) closely follows the linear trend of the HSBs, but the fainter end (\mue{ $>24.5$} \magperarcsec{}) diverges from this trend to form a flattening of \sigmastar{} around $\sim$ $10^{7}$ \msunkpcsq{} for the faintest sources. 

We made an error-weighted linear fit to the full sample (as shown in Fig. \ref{fig:mue_sigmastar}) to obtain a best-fit relation as given in Eq. \ref{eqn:mue_sigmastar},

\begin{equation}\label{eqn:mue_sigmastar}
    \log \Sigma_{\rm star} = (-0.40\pm0.01)\, \bar{\mu}_{\rm e,\, r} + (16.31 \pm 0.13),
\end{equation}
where $\bar{\mu}_{\rm e,\, r}$ and \sigmastar{} are in \magperarcsec{} and \msunkpcsq{} units, respectively. 
Obviously, this relation is determined by the stellar mass-to-light ratio and its eventual dependence on the stellar mass surface density. It is remarkable that we obtain a slope of $-0.4$, as expected if the mass-to-light ratio does not depend on the stellar mass surface density. Our best-fit line lies very close to a constant mass-to-light ratio of 1 M$_{\odot}$/L$_{\odot}$ (see Fig. \ref{fig:mue_sigmastar}).
The majority of our sample is within the $3\sigma$ confidence level of the best-fit line (grey shaded region in Fig. \ref{fig:mue_sigmastar}), except for about 2.5\% of the sample (38 galaxies, among which 36 are LSBs and 2 are HSBs) that lies outside the $3\sigma$ range of the best-fit. These outliers are mostly LSBs with a high stellar mass surface density. This indicates a higher mass-to-light ratio for these galaxies. Using the \textit{r}-band luminosities and the stellar masses of our sample, we estimated that the outliers have a median mass-to-light ratio ($M/L_{r}$) of 3.4 M$_{\odot}$/L$_{\odot}$, compared to 1.1 M$_{\odot}$/L$_{\odot}$ for the full sample, making them distinct outliers. 

Since the definition of our outliers given in Fig. \ref{fig:mue_sigmastar} depends on the choice of the degree of the fit, we also performed a test with a polynomial fit of order 2. We found that the polynomial fit provides a better fit with smaller residuals than the linear fit and reduces the number of outliers from 38 to 11. However, such a fit can also be affected by any incompleteness at the low surface brightness range. Moreover, in the polynomial fit, we lose an important piece of information that we have in the linear fit. The linear fit reproduces very well the trend in the HSB regime, and the outliers in the LSB regime are clearly a population of galaxies that are distinct from their HSB counterparts, as they lie in a range of high fiducial $M/L$ ratio. This is a very distinct behavior, and we are interested in studying those cases. Therefore, from hereupon, we adopt the linear fit as given in Eq. \ref{eqn:mue_sigmastar} and the 38 outliers obtained from it.
%

\subsection{Specific star formation rate}

\begin{figure}
    \centering
    \includegraphics[width=\hsize]{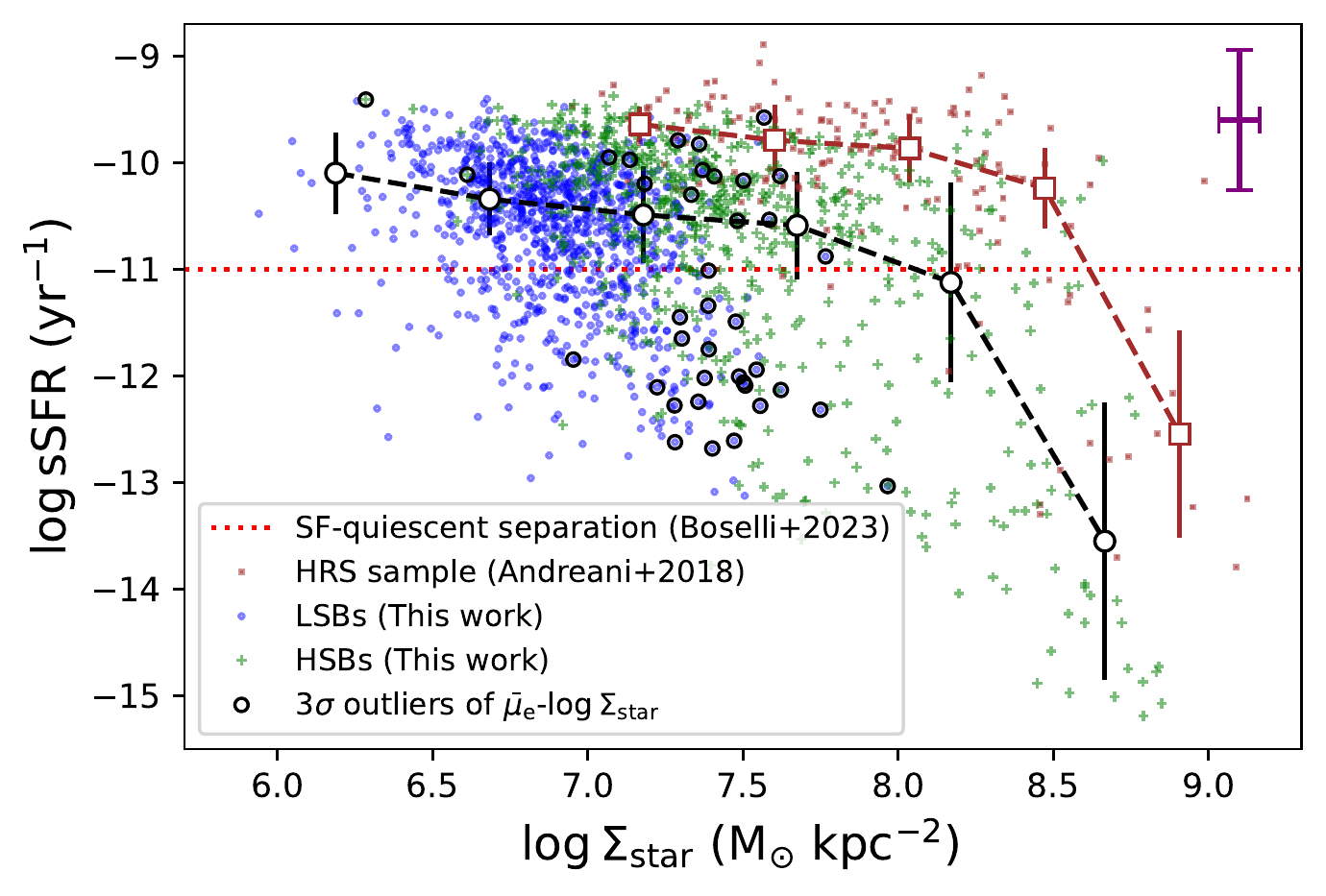}
    \caption{Relationship between the specific star formation rate and stellar mass surface density. The symbols are the same as in Fig. \ref{fig:mue_sigmastar}. The red dashed line marks the separation of star-forming and quiescent galaxies \citep{boselli2023}. The mean uncertainty of the sample obtained by propagating errors on individual measurements is given as the magenta errorbar on the top right corner.}
    \label{fig:ssfr_sigmastar}
\end{figure}

Figure \ref{fig:ssfr_sigmastar} shows the dependence of the specific star formation rate (sSFR) of our sample as a function of the stellar mass surface density. Majority of our sample ($\sim$73\%) are star-forming galaxies with $\rm sSFR > 10^{-11}$ yr$^{-1}$ \citep{boselli2023}. We can see that for the star-forming galaxies, the sSFR is mostly flat with respect to the stellar mass surface density, but with a slight indication of a decrease in sSFR from the low to the high stellar mass surface density until \sigmastar{} $\sim 10^8$ \msunkpcsq{}. Beyond this value, the sSFR shows a sudden decline to reach the population of quiescent galaxies (with a large scatter and big uncertainty in the sSFR of the order of $\sim$1 dex for quiescent galaxies). This trend is similar to what is observed in the HRS sample too, although the HRS sample, on average, has a higher sSFR than our sample. Interestingly, the outliers discussed in Sect. \ref{sect:mue_sigmastar} lie equally along the star-forming and quiescent part of the sample. The LSBs and HSBs, on average, have very similar sSFR values (median $\log\,\rm sSFR$ of $-10.5$ yr$^{-1}$ for the LSBs and $-10.4$ yr$^{-1}$ for the HSBs), in comparison to the slightly higher sSFR of the HRS galaxies (median $\log\,\rm sSFR$ of $-9.9$ yr$^{-1}$). Our sample, therefore, brings an important extension of the sSFR-\sigmastar{} relation in the regime of low surface brightness galaxies.

\subsection{Specific infrared luminosity}

Estimating the dust mass of galaxies requires a proper constraint on the peak of the FIR emission. However, considering the data we have for our sample, it is not possible to determine the dust masses. Therefore, we use the total IR luminosity of our sample obtained from the SED fitting\footnote{The \lir{} values from CIGALE were computed by integrating the full dust emission model (shown as the red solid curves in Fig. \ref{fig:sed_bestmodel}) over an arbitrarily large wavelength range used in the modeling. This should, in practice, give very similar values as the \lir{} commonly estimated in the literature within the wavelength range of 8-1000 $\mu$m.} discussed in Sect. \ref{section:sed_fitting} as a proxy for the dust mass \citep[e.g.,][]{dacunha2010,Orellana2017}. Similarly, the ratio of the \lir{} and stellar mass (L$_{\rm IR}$/M$_{\rm star}$, which we term here as the specific infrared luminosity or the sLIR) is used to probe the specific dust mass (M$_{\rm dust}$/M$_{\rm star}$) of our sample. Specific dust mass of galaxies is an important measure of dust production \citep[e.g.,][]{Cortese2012b,Casasola2020}, as well as dust destruction processes and dust re-formation mechanisms \citep[e.g.,][]{Casasola2020,Donevski2020}.

\begin{figure}
    \centering
    \includegraphics[width=\hsize]{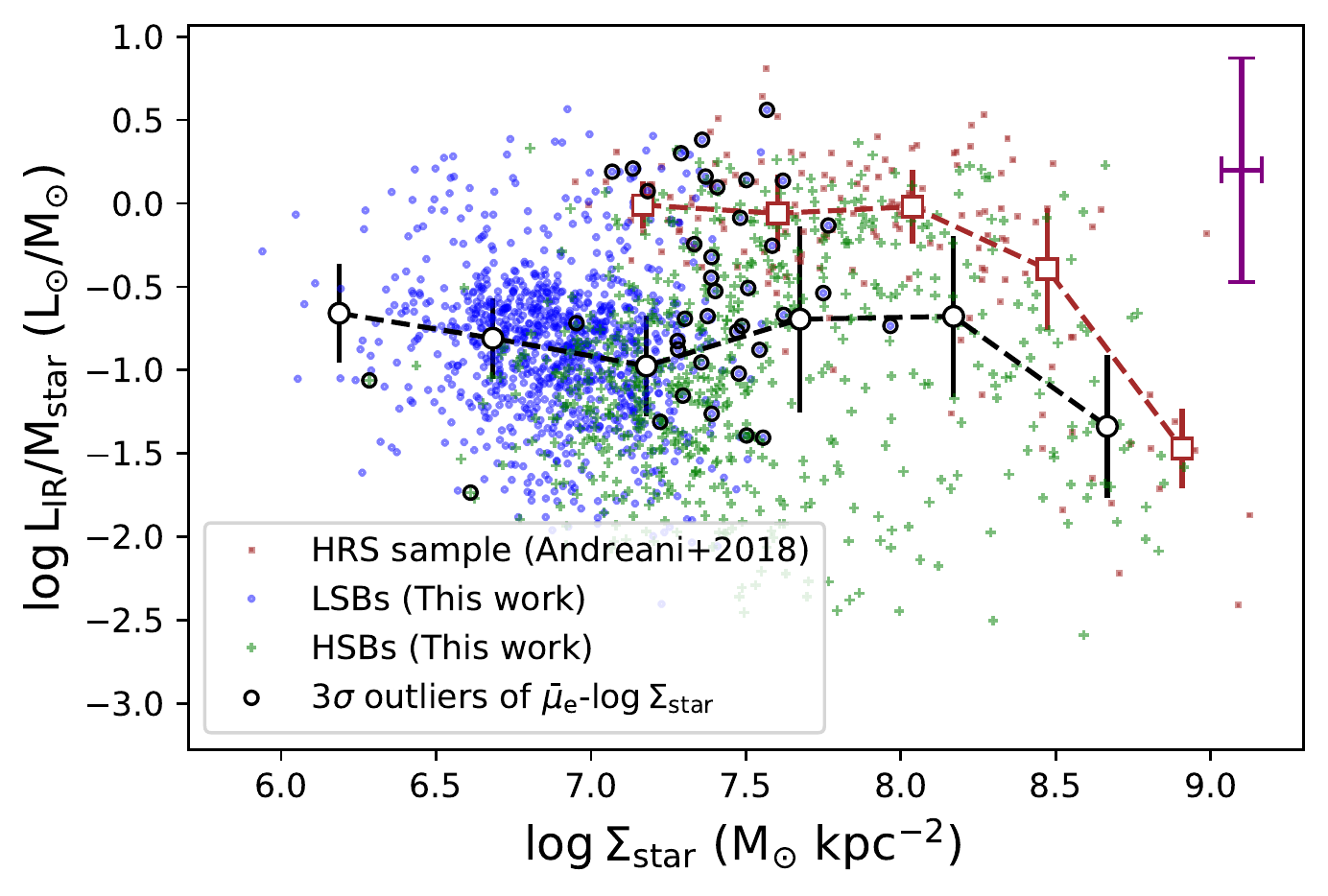}
    \caption{Specific infrared luminosity (\lir{} per unit stellar mass) as a function of the stellar mass surface density. The symbols are the same as the previous figures. The mean uncertainty of the sample obtained by propagating errors on individual measurements is given as the magenta errorbar on the top right corner.}
    \label{fig:specific_LIR_sigmastar}
\end{figure}

Figure \ref{fig:specific_LIR_sigmastar} shows the variation of the specific infrared luminosity as a function of the stellar mass surface density. 
 At the brightest end of Fig. \ref{fig:specific_LIR_sigmastar}, the sLIR rises steeply with decreasing stellar mass surface density until \sigmastar{} $\sim$$10^8$ \msunkpcsq{}, which is similar to the trend seen in Fig. \ref{fig:ssfr_sigmastar} for the sSFR of quiescent galaxies. This steep rise is observed for the HRS sample too. Below \sigmastar{} $\sim$$10^8$ \msunkpcsq{}, the sLIR remains mostly flat towards lower stellar mass surface densities, as also seen in the HRS sample, which however, lies along the higher sLIR part of the sample. Therefore our sample allowed us to explore the trend of increasing specific dust content with decreasing the stellar density, at lower densities than what was found in HRS. We find that dust emission is present at low densities, but with saturation in the specific dust content rather than an increase.
Moreover, similar to what was observed with the sSFR, both the LSBs and HSBs in our sample, on average, have comparable sLIR values of $10^{-0.9}$ L$_{\odot}$/M$_{\odot}$ and $10^{-1.1}$ L$_{\odot}$/M$_{\odot}$, respectively. The HRS, on the other hand, lies along the higher sLIR tail of the distribution with a median value of $10^{-0.2}$ L$_{\odot}$/M$_{\odot}$. A fraction of our HSBs also has sLIR similar to what is found in HRS. The outliers from the \mue{}-\sigmastar{} relation occupies the transition region of low to high sLIR, with many having high sLIR values as the HRS sample. Therefore, from the distribution given in Fig. \ref{fig:specific_LIR_sigmastar}, we can infer that LSBs have similar sLIR values to that of HSBs, although they have a lower absolute \lir{}. Moreover, since we observe a similar trend in sLIR and sSFR, both these quantities might be related. However, it is hard to disentangle them based on star formation activity and dust emission since we do not know much about the infrared properties of such galaxies.




\subsection{Dust attenuation} \label{sect:dust_attenuation}

Figure \ref{fig:av_sigmastar} shows the \textit{V}-band dust attenuation (\Av{}) of the sample with respect to the stellar mass surface density. The majority of our sample ($\sim$ 60\%) have a low attenuation with \Av{ $<0.1$} mag. For the highest \sigmastar{} sources, which are well constrained with small uncertainties, we observe a higher attenuation, but with a large scatter. For the fainter galaxies, the attenuation steeply decreases to reach an almost negligible value close to zero. However, the uncertainties associated with the \Av{} estimates of many of these faint sources are typically large. For instance, the galaxies with \sigmastar{} $<10^7$ \msunkpcsq{} and with \Av{ $>0.5$} have an uncertainty in \Av{} estimation of the order of 0.4 mag, making it hard to draw conclusions on them. Nevertheless, we still observe several faint galaxies with significant attenuation and small uncertainties. The 3$\sigma$ outliers of the \mue{}-\sigmastar{} relation discussed in Sect. \ref{sect:mue_sigmastar} (38 galaxies) are among them which appears to be an interesting group in terms of attenuation. From Fig. \ref{fig:av_sigmastar}, we see that about 60\% of the outliers (23 out of 38 galaxies) have a large attenuation with \Av{ $>0.5$} mag and a mean value of $0.8\pm0.2$ mag. Moreover, several of these outliers also have detection in the IRAC bands, similar to the IRAC-detected LSBs from \citet{hinz2007}. However, none of the outliers have any detection in the MIR or FIR range.

Following \citet{boselli2023}, who derived a relation between attenuation and stellar mass, we did an error-weighted fit to our data to find a similar relation between \Av{} and \sigmastar{} of our sample as given in Eq. \ref{eqn:av_sigmastar}:

\begin{figure}
    \centering
    \includegraphics[width=\hsize]{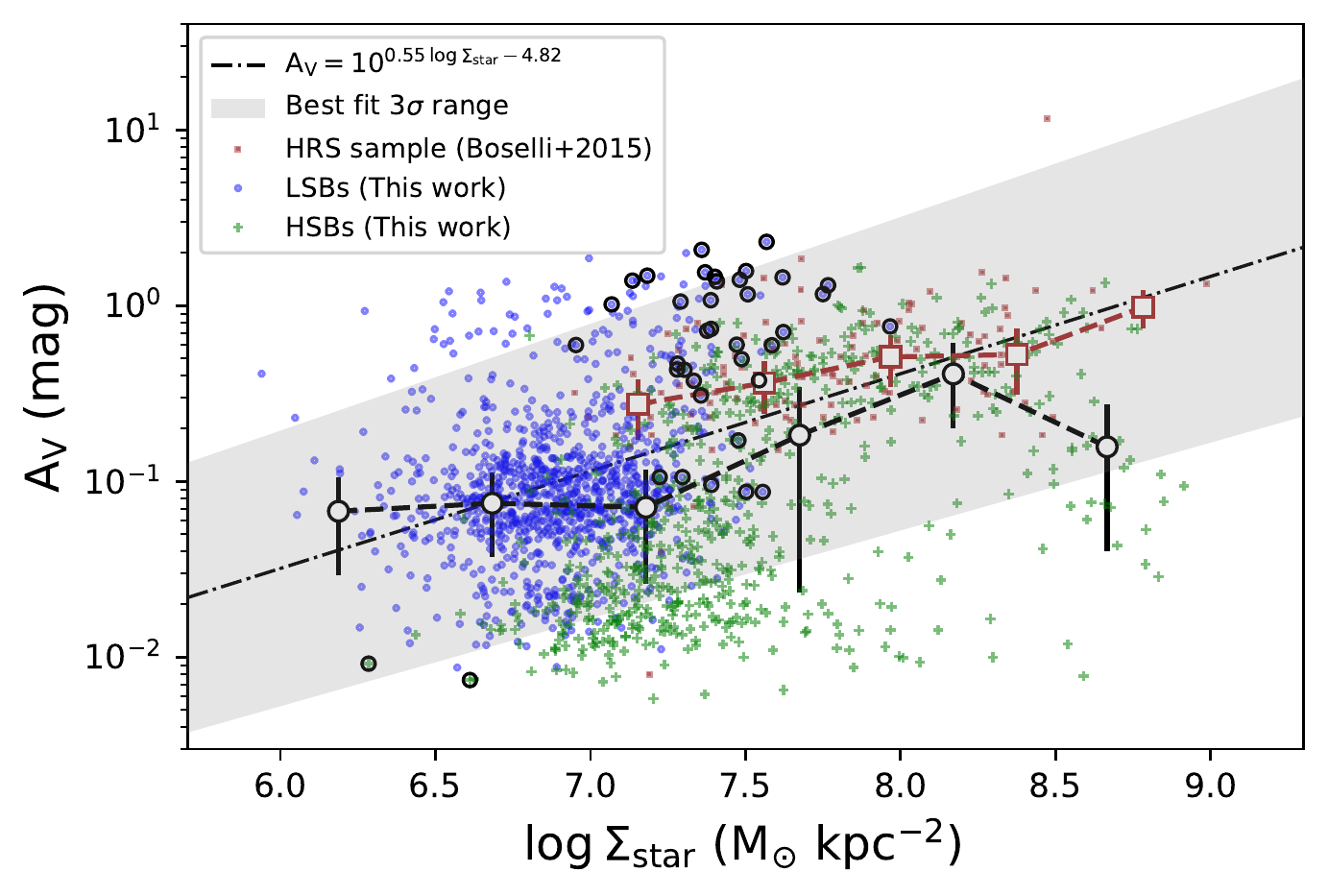}
    \caption{\textit{V}-band attenuation of the sample as a function of the stellar mass surface density. The black dot-dashed line is the best-fit line for our sample, and the gray shaded region is its corresponding $3\sigma$ uncertainty. The black open circles around some sources are the 3$\sigma$ outliers of the \mue{}-\sigmastar{} relation as shown in Fig. \ref{fig:mue_sigmastar}, and discussed in Sect. \ref{sect:mue_sigmastar}. The \Av{} values of our sample have a mean uncertainty of 0.12 mag.}
    \label{fig:av_sigmastar}
\end{figure}

\begin{equation}\label{eqn:av_sigmastar}
    A_{V} \, \rm (mag) = 10^{(0.55\pm0.02)\, \log \Sigma_{\rm star} - (4.82\pm0.15)}.
\end{equation}

Our best-fit relation also follows a trend where \Av{ $<0.1$} mag for the faint galaxies until \sigmastar{} $\sim$$10^7$ \msunkpcsq{}, beyond which we see a steep rise in \Av{} for the brighter galaxies, with a large scatter (note that the scatter shown in Fig. \ref{fig:av_sigmastar} is in logarithmic scale). The HRS sample shows a similar trend in attenuation with the stellar mass surface density, although in general, it has larger \Av{} than our sample for the same \sigmastar{}, but consistent with the large scatter seen in this range. Note that only the late-type galaxies from the HRS sample have available measurements in \Av{} (see Sect. \ref{sect:comparison_sample}). This explains the lack of high \sigmastar{} HRS galaxies with attenuation close to zero, as observed in our sample.

We also found that the steep rise of \Av{} in Fig. \ref{fig:av_sigmastar} is largely driven by the galaxies at \sigmastar{$\,> 10^{8}$} \msunkpcsq{}, that are dominated by more massive HSBs (we do not have any LSBs beyond this \sigmastar{} value). So the trend in \Av{} we see here is also linked to its dependence on the stellar mass, which is well known \citep[e.g.,][]{Bogdanoska2020,Riccio2021,boselli2023}. However, in the range of \sigmastar{$\,< 10^{8}$} \msunkpcsq{}, we have a large overlap with the LSBs and HSBs of stellar masses mostly in the range of $10^8-10^9$ \msun{}. They are both consistent with low attenuation, except for the LSB outliers that remain as a peculiar population with high attenuation.


\section{Discussions}\label{sect:discussion}

\subsection{Are low surface brightness galaxies dust-free?}


%
%
The results given in Sect. \ref{sect:dust_attenuation} show that the majority of the LSBs in our sample (\mue{ $>23$} \magperarcsec{} or approximately \sigmastar{} $<10^7$ \msunkpcsq{})\footnote{A stellar mass surface density of $10^7$ \msunkpcsq{} corresponds to an average \textit{r}-band surface brightness (\mue{}) of 23.2 \magperarcsec{}, based on the Eqn. \ref{eqn:mue_sigmastar}.} have a very low amount of dust attenuation. Among the LSBs (1003 out of 1631 galaxies), about 80\% have a negligible attenuation with an \Av{ $<0.2$} mag, and a median attenuation of $\sim$0.09 mag. This is consistent with few other observations of LSBs from the literature where extreme LSBs like the ultra-diffuse galaxies (UDGs) were found to have a very low attenuation with a median \Av{} of $\sim$0.1 mag \citep{pandya2018, Barbosa2020, Buzzo2022}. However, \citet{Liang2010} found a median \Av{} of 0.46 mag for their sample of LSBs from the SDSS survey. Such a higher attenuation in their LSB sample could be attributed to the fact that the LSBs from \citet{Liang2010} where massive galaxies with a median stellar mass of $10^{9.5}$ \msun{}, in comparison to our low-mass LSBs with a median stellar mass of $10^{8.3}$ \msun{}. Moreover, the \Av{} values from \citet{Liang2010} were computed from the Balmer decrement, without applying a correction for the differential attenuation of nebular lines and the continuum as shown in Table \ref{table:cigale_params}. Applying such a correction will reduce their median \Av{} to $\sim$0.2 mag, which is close to the values we observe for our sample of LSBs.
Only about 4\% of the LSBs in our sample (2.5\% of the total sample) have a significant attenuation with \Av{ $>0.5$} mag. These are 3$\sigma$ outliers from the \mue{}-\sigmastar{} relation as shown Fig. \ref{fig:mue_sigmastar} and Fig. \ref{fig:av_sigmastar}. This could indicate that a fraction of low surface brightness galaxies with high stellar mass-to-light ratios ($M/L_{r} > 3$ M$_{\odot}$/L$_{\odot}$) seem to have a higher attenuation. 

\begin{figure}
    \centering
    \includegraphics[width=\hsize]{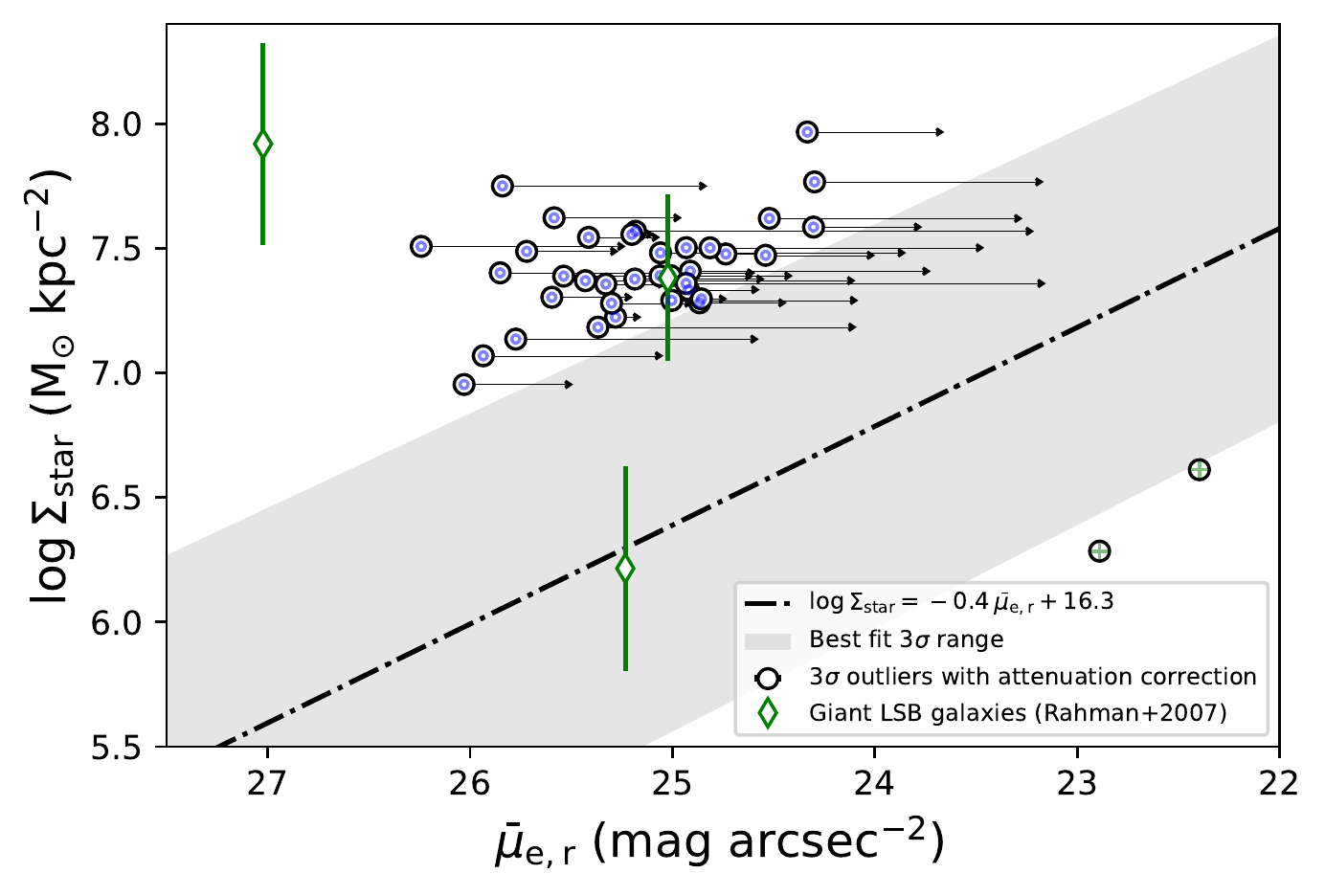}
    \caption{The $3\sigma$ outliers of the surface brightness-stellar mass surface density linear relation discussed in Sect. \ref{sect:mue_sigmastar}. The black arrows indicate the directions/positional changes on this plane, and how the outliers (open circles) will move after the correction for the \textit{r}-band attenuation.
    The green diamond symbols mark the location of three giant LSB galaxies (Malin 1, UGC 9024, and UGC 6614, from the left to right, respectively) from the literature \citep{rahman2007}. The black dash-dotted and the grey-shaded region is the best-fit line and its $3\sigma$ confidence range as described in Sect. \ref{sect:mue_sigmastar}.}
    \label{fig:outliers_attenuation_corrected}
\end{figure}

We also looked into how much the attenuation affects the position of the outliers in the \mue{}-\sigmastar{} relation. For this purpose, we applied a correction for the observed surface brightness of the outliers using the estimated \Av{} values. We converted the \textit{V}-band attenuation to the attenuation in the HSC \textit{r}-band using the \citet{Calzetti2000} attenuation law, before correcting for the \textit{r}-band surface brightness. Figure \ref{fig:outliers_attenuation_corrected} shows the change in the position of the outliers after the attenuation correction. All the outlier LSBs still remain LSBs with \mue{ $<23$} \magperarcsec{}. However, we can see that about 50\% of them move into the $3\sigma$ confidence range of the \mue{}-\sigmastar{} relation after the attenuation is corrected. This indicates that the effect of attenuation plays a significant role in making a fraction of LSBs appear fainter in the observations. However, attenuation alone cannot explain all the outliers in our \mue{}-\sigmastar{} relation.


Giant low surface brightness galaxies (GLSBs) are another interesting and extreme class of objects among LSBs \citep[e.g.,][]{Sprayberry1995,Hagen2016,junais2020}. The infrared properties of three GLSBs (Malin 1, UGC 6614, and UGC 9024) were explored by \citet{rahman2007} using \textit{Spitzer} observations. All of them were undetected at MIR and FIR wavelengths allowing only to obtain upper limits in their infrared properties. Figure \ref{fig:outliers_attenuation_corrected} shows a comparison of their stellar mass surface density\footnote{The stellar masses and sizes of the GLSBs were taken from \citet{Comparat2017} and \citet{rahman2007}, respectively, to estimate their stellar mass surface densities.} and surface brightness\footnote{The \mue{} values of the GLSBs were estimated by using the \textit{B}-band central surface brightness ($\mu_{0, B}$) values from \citet{rahman2007}. The $\mu_{0, B}$ values were converted to the \textit{r}-band \mue{} assuming a constant S\'ersic index $n=1$ \citep{Graham2005}, and a constant $B-r$ color of 0.6 mag.} as compared to our sample.
%
%
We can see that two out of the three GLSBs (Malin 1 and UGC 6614) are 3$\sigma$ outliers from the \mue{}-\sigmastar{} relation. Moreover, their sSFR and sLIR are also well consistent with our sample (based on \citealt{rahman2007}, the three GLSBs have an sSFR of $10^{-10.8}$, $10^{-10.2}$ and $10^{-10.4}$ yr$^{-1}$, and sLIR of $10^{-0.9}$, $10^{-0.4}$, $10^{-0.5}$ L$_{\odot}$/M$_{\odot}$, respectively). Therefore, from Fig. \ref{fig:outliers_attenuation_corrected}, it is likely that these GLSBs also have a significant dust attenuation, similar to the outliers we observe in our sample, although their previous infrared observations do not provide any estimate of attenuation. 

Apart from the observational data on GLSBs, \citet{kulier2020} provided some estimates on the dust attenuation of GLSBs from the EAGLE simulations (see their Fig. A2). They obtained an average \Av{} of 0.15 mag for their simulated GLSB sample, with a range of values extending from \Av{ $=0.4$} mag for the brighter sources (\mue{$\sim$23} \magperarcsec{}) to \Av{ $=0.05$} mag for the faintest ones (\mue{$\sim$26} \magperarcsec{}). Therefore, comparing our results with observations and simulations allows expecting the presence of some detectable dust attenuation in GLSBs as well.



\subsection{Possible caveats}

The analyses presented in this work could be affected by several caveats. Firstly, since we attempt to study the optical as well as infrared properties of our sample (surface brightness, radius, stellar mass, SFR, total infrared luminosity, and dust attenuation), it requires extensive multi-wavelength data coverage in the UV to FIR range. However, as noted in Sect. \ref{sect:crossmatch-catalogs}, for $\sim$85\% of our sample, only the deep $5\sigma$ upper limits can be provided in the MIR to FIR regime (from 7 $\mu$m to 500 $\mu$m wavelength range). In those cases, these detection limits are used to put constraints on the infrared emission of the SED. Such an approach can introduce significant uncertainties in the estimated infrared properties of our sample (especially in the \lir{} and \Av{}). We performed several tests to quantify and minimize 
 the effect of such uncertainties on our results (see Sect. \ref{sect:robustness} for more details on the robustness of the estimated parameters). Additionally, our sample selection with the requirement to have a \textit{u}-band detection, is also aimed to minimize such uncertainties. The \textit{u}-band, being close to the UV part of the spectrum, is more sensitive to the effects of dust attenuation and thereby the re-emission in the infrared. 

Another potential uncertainty in our results can arise from a possible redshift dependence of the quantities. However, considering the very narrow range of redshift used in this work ($z<0.1$), and since we used redshift-independent quantities for this work, we do not expect to have an impact of such an effect. We verified that there are no significant variations in our sample with the redshift, and our results will remain unchanged. However, the accuracy of the photometric redshift estimates used in this work from \cite{huang2021} can be yet another source of uncertainty. Considering the very faint nature of the majority of our sample, it is not feasible to obtain spectroscopic redshifts for all of them (all the galaxies with spectroscopic redshifts in our sample are HSBs, as shown in Table \ref{appendix:table_estimated_properties}). Also, as discussed in Sect. \ref{sample_selection}, in general, the photometric redshift estimates we used have a higher accuracy and a lower catastrophic outlier rate. The presence of the \textit{u}-band also significantly improves the photometric redshift estimates, as noted by \citet{huang2021}. Nevertheless, we made an estimate of the effect of the photometric redshift uncertainty on our measured physical quantities. For a typical redshift uncertainty of $\sigma_{z_{\rm p}}=0.06$ for our sample (as discussed in Sect. \ref{sample_selection}), we found that, on average, the \mstar{} and \Re{} changes by a large factor (0.47 dex and 0.21 dex, respectively). However, since we compute the \sigmastar{} as a ratio of \mstar{} and \Re{} as given in Eq. \ref{eqn:sigma_star}, they cancel each other for the \sigmastar{} to have only a 0.04 dex difference with the change in redshift, making \sigmastar{} an almost redshift-independent quantity. In the case of the sSFR, sLIR and \Av{} also we see only a negligible difference (0.13 dex, 0.04 dex, and 0 dex, respectively).

Therefore, considering all the above potential caveats, we conclude that our estimates are still robust within the uncertainties discussed. The approach we used in this work will be useful for constraining the physical quantities of LSBs, especially with the upcoming surveys like LSST that will observe thousands of them in the \textit{ugrizy}-bands, with only limited multi-wavelength counterparts.

\section{Conclusions}\label{sect:conclusion}

We present an optically selected sample of 1631 galaxies at $z<0.1$ from the North Ecliptic Pole Wide field. We cross-matched this sample with several multi-wavelength sets of available data ranging from UV to FIR, and did a SED fitting procedure to obtain key physical parameters like the stellar mass, SFR, \lir{} and \Av{}. We also extracted radial surface brightness profiles for the sample and estimated their average optical surface brightness and sizes.
Our main results can be summarised as follows:

\begin{itemize}

    \item Using the measured average \textit{r}-band surface brightness (\mue{}), our sample consists of 1003 low surface brightness galaxies (LSBs; \mue{ $>23$} \magperarcsec{}) and 628 high surface brightness galaxies (HSBs; \mue{ $\leq23$} \magperarcsec{}).
    
    \item The LSBs have a median stellar mass, surface brightness, and effective radius of $10^{8.3}$ \msun{}, 23.8 \magperarcsec{} and 1.9 kpc, respectively. For the HSBs, the corresponding median values are $10^{8.8}$ \msun{}, 22.2 \magperarcsec{} and 2.2 kpc. Similarly, the LSBs have a median SFR and \lir{} of $10^{-2.2}$ \msun{} yr$^{-1}$ and $10^{7.4}$ \lsun{}, in comparison to $10^{-1.6}$ \msun{} yr$^{-1}$ and $10^{7.7}$ \lsun{} for the HSBs. For both the LSBs and HSBs, we found a median \Av{} of 0.1 mag.
    
    \item A comparison of the surface brightness (\mue{}) as a function of the stellar mass surface density (\sigmastar{}) showed that our sample follows the linear trend for the HSBs, which is consistent with the HRS sample from the literature. However, for the LSBs, we observe several outliers from the linear \mue{}-\sigmastar{} relation, indicating a higher mass-to-light ratio for them. Most of these outliers also have a high dust attenuation.
    
    \item We analyzed the variation of the specific star formation rate (sSFR) and specific infrared luminosity (sLIR) of our sample with respect to their stellar mass surface density. Among the star-forming galaxies (sSFR $>10^{-11}$ yr$^{-1}$), the sSFR is mostly flat with respect to the change in stellar mass surface density, but with a slight indication of an increase in sSFR for the lowest \sigmastar{} galaxies. The sSFR steeply declines for the highest \sigmastar{} sources that are quiescent. A similar trend is observed for the sLIR too. We found that both the LSBs and HSBs in our sample have a comparable average sSFR and sLIR. The HRS sample, in general, lies along the higher sSFR and sLIR regime compared to our sample but they are consistent and agree within the scatter we observe.

    \item The change in dust attenuation (\Av{}) with the stellar mass surface density of our sample shows that galaxies with a higher \sigmastar{} have a larger \Av{} and scatter, contrary to the flat/decreasing trend observed for the specific dust luminosity. The dust attenuation steeply declines and becomes close to zero for the majority of LSBs.
    However, in about 4\% of the LSBs that are outliers, we observe a significant attenuation with a mean \Av{} of 0.8 mag, showing that not all the LSBs are dust poor. Moreover, the extreme giant LSBs in the literature also show some similarities to these outlier LSBs, indicating the presence of more dust content in them than previously thought.    
\end{itemize}

This work provides measurements that can be further tested using current/upcoming observations from LSST and JWST, where a large number LSBs and HSBs will be observed at unprecedented depth. LSST will provide deep optical imaging data over large areas of the sky, allowing for a detailed study of the statistical properties of galaxies, including LSBs. On the other hand, JWST's high sensitivity and resolution imaging in the near-infrared (NIR) and mid-infrared (MIR) regimes, as well its spectroscopic capabilities, will enable a comprehensive study of the infrared properties of such galaxies, including their dust content, gas metallicity and star formation activity. The data from these facilities will complement this work to provide a clear picture of the properties of LSBs in the context of galaxy evolution.


\begin{acknowledgements}

J and KM are grateful for support from the Polish National Science Centre via grant UMO-2018/30/E/ST9/00082. W.J.P. has been supported by the Polish National Science Center project UMO-2020/37/B/ST9/00466 and by the Foundation for Polish Science (FNP). J.K. acknowledges support from NSF through grants AST-1812847 and AST-2006600. M.R. acknowledges support from the Narodowe Centrum Nauki (UMO-2020/38/E/ST9/00077). M.B. gratefully acknowledges support by the ANID BASAL project FB210003 and from the FONDECYT regular grant 1211000. D.D acknowledges support from the National Science Center (NCN) grant SONATA (UMO-2020/39/D/ST9/00720). D.D also acknowledges support from the SISSA visiting research programme. A.P. acknowledges the Polish National Science Centre grant UMO-2018/30/M/ST9/00757 and the Polish Ministry of Science and Higher Education grant DIR/WK/2018/12.

\end{acknowledgements}

\bibliographystyle{aa}
\bibliography{bibliography}

\begin{appendix}
\onecolumn


\section{SED fitting robustness}
\subsection{Mock analysis}\label{appendix:sect_mock_analysis}

\begin{figure*}
    \centering
    \includegraphics[width=\hsize]{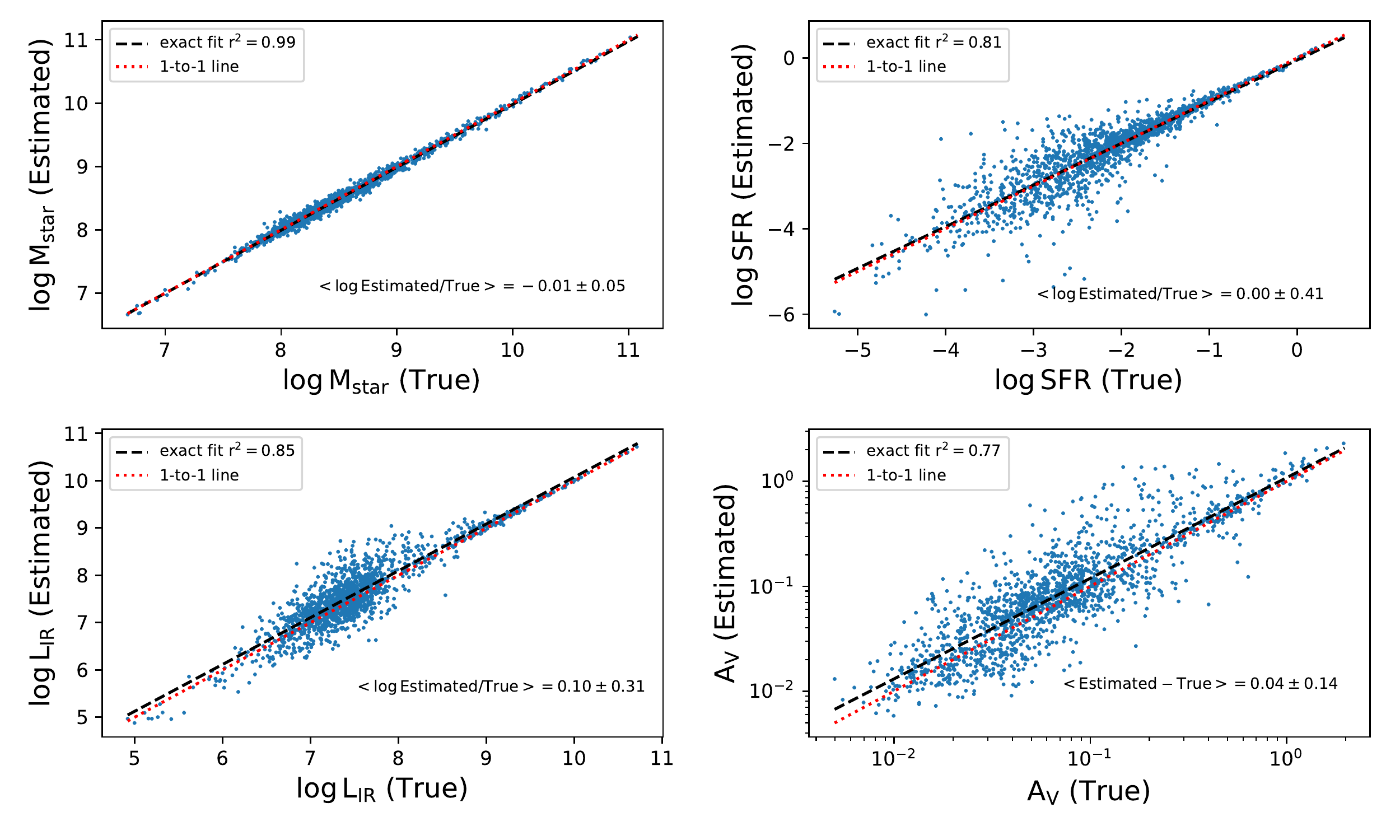}
    \caption{Mock analysis performed using \cigale{} to compare the ``true'' values from the mock catalog with the Bayesian estimated values. The black dashed lines show the linear regression fit and the corresponding regression coefficients ($r^{2}$) are marked in each panel. The red dotted line is the one-to-one relation. The mean difference and scatter between the estimated and true values are marked in each panel.}
    \label{appendix:fig_mocks}
\end{figure*}

\subsection{Comparison of fits with and without FIR data}\label{appendix:sect_fir_ugrizy_comparison}

\begin{figure*}
    \centering
    \includegraphics[width=\hsize]{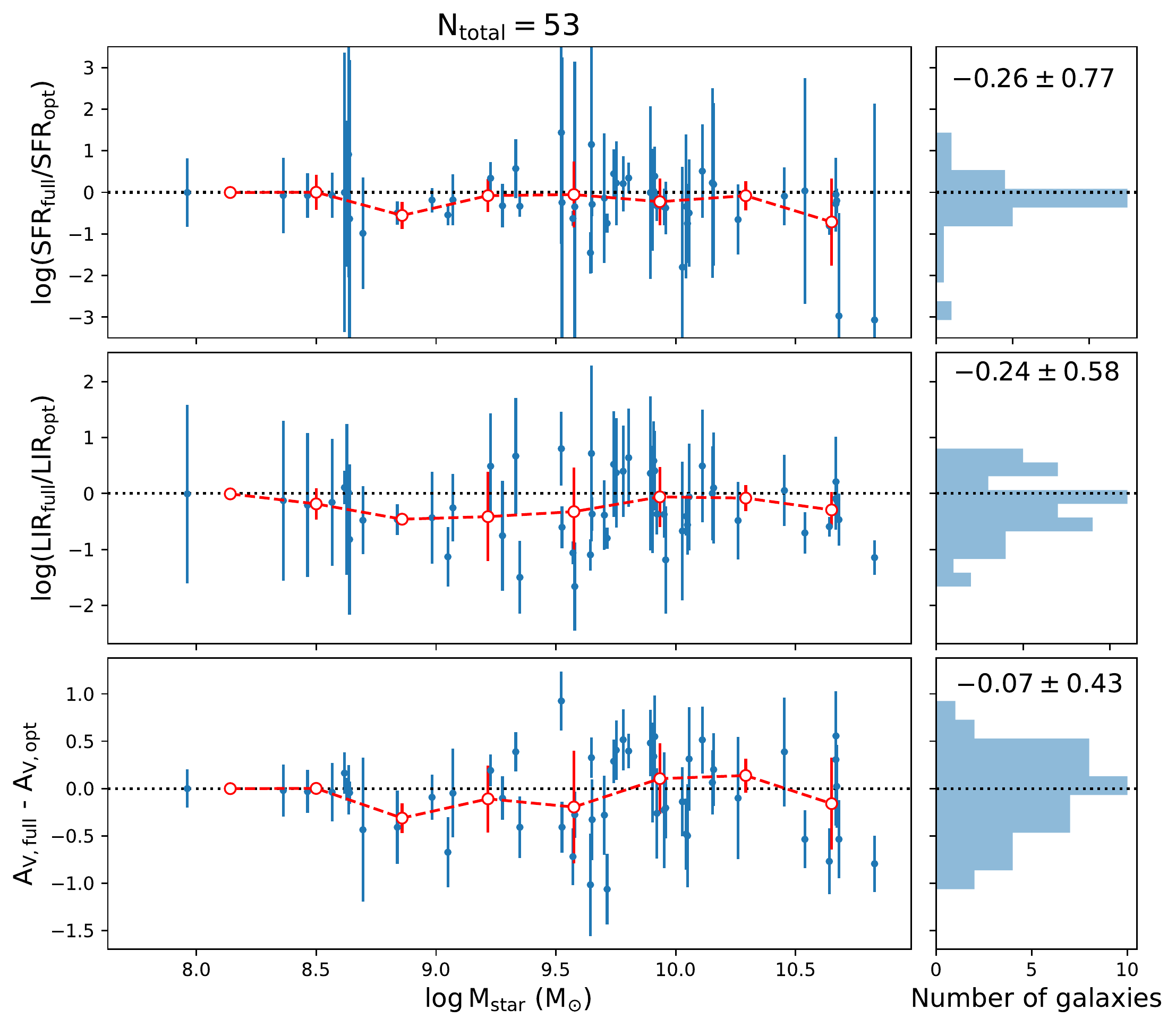}
    \caption{Comparison of the SED fitting results for the 53 FIR detected galaxies using their full photometry from UV to FIR with respect to a fitting using only the optical \textit{ugrizy}-bands. From the top to the bottom panels shows the difference in SFR, \lir{} and \Av{} estimated from the two fits, as a function of the stellar mass of the galaxies. The red dashed line marks the mean and scatter along different stellar mass bins. The histograms beside each panel give the overall distribution of each quantity, with their mean and scatter marked on top of each histogram. For the stellar mass estimates between the two fits, there is only a minor difference as expected, with a mean difference of $-0.09\pm0.13$ dex.}
    \label{appendix:fig_fir_ugrizy}
\end{figure*}

\section{Data table with the physical properties of the sample}\label{appendix:sect_data_table}

\begin{table*}
\centering
\caption{Estimated properties of the sample. The complete table will be available on CDS.}\label{appendix:table_estimated_properties}
\fontsize{8.5}{8.5pt}
\begin{tabular}{|*{10}{c|}}
\hline
ID & R.A & Dec. & \textit{z} & \mue{} & \Re{} & $\log$ \mstar{} & $\log$ SFR & $\log$ \lir{} & \Av{} \\
 & (deg) & (deg) &  & (mag/\arcsec$^{2}$) & (kpc) & (\msun{}) & (\msun{} yr$^{-1}$) & (L$_{\odot}$) & (mag) \\
(1) & (2) & (3) & (4) & (5) & (6) & (7) & (8) & (9) & (10) \\
\hline
79666648293863491 & 267.7464 & 66.4367 & 0.05 & 21.5 & 4.0 & $9.11\pm0.04$ & $-0.80\pm0.06$ & $7.72\pm0.50$ & $0.01\pm0.01$ \\
79666652588826778 & 267.7954 & 66.5453 & 0.09 & 21.4 & 2.6 & $9.15\pm0.03$ & $-0.46\pm0.04$ & $9.05\pm0.08$ & $0.14\pm0.02$ \\
79666648293856204 & 267.8829 & 66.3565 & 0.08 & 23.9 & 1.6 & $8.25\pm0.11$ & $-3.06\pm1.40$ & $7.23\pm1.04$ & $0.12\pm0.20$ \\
79666656883786450 & 267.9090 & 66.7056 & 0.02 & 20.4 & 1.3 & $9.07\pm0.03$ & $-1.10\pm0.05$ & $8.93\pm0.02$ & $0.42\pm0.01$ \\
79666643998886722 & 267.9358 & 66.1153 & 0.10 & 23.4 & 1.5 & $8.21\pm0.06$ & $-1.69\pm0.16$ & $7.58\pm0.82$ & $0.06\pm0.10$ \\
79666506559930664 & 268.0367 & 66.1148 & 0.10 & 23.4 & 3.4 & $8.64\pm0.04$ & $-1.38\pm0.17$ & $7.21\pm0.65$ & $0.01\pm0.02$ \\
79666519444831698 & 268.0397 & 66.6789 & $0.09^{sp}$ & 20.7 & 3.2 & $9.80\pm0.03$ & $0.04\pm0.02$ & $9.98\pm0.02$ & $0.51\pm0.01$ \\
80093649647458461 & 268.0701 & 66.9625 & 0.09 & 23.4 & 4.3 & $8.96\pm0.07$ & $-1.66\pm0.37$ & $7.97\pm0.66$ & $0.08\pm0.10$ \\
79666510854915341 & 268.1099 & 66.4497 & 0.07 & 19.6 & 2.6 & $9.91\pm0.03$ & $0.18\pm0.02$ & $10.07\pm0.02$ & $0.42\pm0.00$ \\
79666515149883548 & 268.1283 & 66.6581 & 0.09 & 21.5 & 2.6 & $9.32\pm0.04$ & $-0.48\pm0.04$ & $9.47\pm0.02$ & $0.52\pm0.03$ \\
79666515149871529 & 268.1389 & 66.5322 & 0.07 & 23.7 & 1.8 & $7.99\pm0.05$ & $-2.09\pm0.19$ & $6.84\pm0.82$ & $0.02\pm0.04$ \\
80093649647472232 & 268.1434 & 67.0023 & 0.08 & 23.7 & 2.6 & $8.23\pm0.06$ & $-2.99\pm0.98$ & $6.80\pm0.83$ & $0.03\pm0.05$ \\
79666515149871597 & 268.1557 & 66.5445 & 0.05 & 21.2 & 1.6 & $8.84\pm0.02$ & $-1.47\pm0.11$ & $7.03\pm0.50$ & $0.01\pm0.01$ \\
79666506559930076 & 268.1632 & 66.1190 & 0.09 & 23.8 & 2.2 & $8.79\pm0.07$ & $-3.28\pm1.00$ & $6.89\pm0.89$ & $0.03\pm0.05$ \\
79666519444845935 & 268.1921 & 66.7848 & 0.10 & 22.2 & 2.9 & $8.94\pm0.03$ & $-1.05\pm0.13$ & $7.63\pm0.71$ & $0.01\pm0.02$ \\
79666510854908456 & 268.2256 & 66.3807 & 0.09 & 22.8 & 3.4 & $8.80\pm0.05$ & $-0.91\pm0.14$ & $8.77\pm0.30$ & $0.21\pm0.13$ \\
79666519444846725 & 268.2300 & 66.7940 & 0.08 & 22.0 & 4.0 & $9.19\pm0.03$ & $-0.74\pm0.08$ & $9.00\pm0.07$ & $0.21\pm0.03$ \\
79666515149881249 & 268.2501 & 66.6152 & 0.03 & 22.2 & 2.7 & $8.99\pm0.06$ & $-2.02\pm0.34$ & $7.36\pm0.56$ & $0.04\pm0.05$ \\
79666510854899112 & 268.2621 & 66.3103 & 0.04 & 20.9 & 1.1 & $9.21\pm0.06$ & $-2.93\pm0.72$ & $8.19\pm0.10$ & $0.26\pm0.05$ \\
79666497969997834 & 268.2715 & 65.7351 & 0.08 & 21.3 & 1.0 & $8.53\pm0.04$ & $-1.23\pm0.08$ & $7.52\pm0.66$ & $0.02\pm0.03$ \\
79666493675047662 & 268.2760 & 65.7109 & 0.06 & 23.9 & 1.6 & $8.37\pm0.09$ & $-3.25\pm1.39$ & $7.07\pm1.10$ & $0.08\pm0.16$ \\
79666506559918377 & 268.2993 & 66.2451 & 0.09 & 25.7 & 1.4 & $8.56\pm0.08$ & $-3.45\pm2.61$ & $7.82\pm0.42$ & $0.50\pm0.31$ \\
79666515149885887 & 268.3016 & 66.6377 & 0.02 & 20.8 & 0.6 & $8.63\pm0.06$ & $-2.00\pm0.11$ & $8.27\pm0.02$ & $0.62\pm0.04$ \\
79666506559940790 & 268.3034 & 66.1899 & 0.06 & 22.5 & 3.2 & $9.47\pm0.04$ & $-1.32\pm0.16$ & $8.82\pm0.04$ & $0.34\pm0.04$ \\
79666497970006035 & 268.3108 & 65.7937 & $0.08^{sp}$ & 21.8 & 5.4 & $9.60\pm0.03$ & $-0.43\pm0.06$ & $9.31\pm0.03$ & $0.20\pm0.02$ \\
79666493675038278 & 268.3172 & 65.6209 & 0.06 & 23.5 & 1.1 & $7.98\pm0.09$ & $-3.05\pm0.87$ & $6.93\pm1.08$ & $0.08\pm0.14$ \\
79218472751488326 & 268.3253 & 65.4307 & 0.10 & 22.4 & 2.0 & $8.85\pm0.08$ & $-1.41\pm0.22$ & $8.15\pm0.57$ & $0.10\pm0.12$ \\
79666493675031327 & 268.3349 & 65.5583 & 0.08 & 22.9 & 2.9 & $8.45\pm0.04$ & $-1.28\pm0.09$ & $7.59\pm0.68$ & $0.03\pm0.04$ \\
80093512208515605 & 268.3610 & 67.0108 & $0.09^{sp}$ & 21.4 & 3.2 & $9.43\pm0.04$ & $-1.57\pm0.74$ & $9.06\pm0.06$ & $0.39\pm0.05$ \\
79666493675047664 & 268.3819 & 65.7192 & 0.10 & 24.1 & 2.0 & $8.01\pm0.06$ & $-1.94\pm0.20$ & $7.48\pm0.80$ & $0.08\pm0.13$ \\
79666493675043780 & 268.3893 & 65.6726 & 0.09 & 23.4 & 1.6 & $8.23\pm0.07$ & $-1.97\pm0.30$ & $7.52\pm0.92$ & $0.08\pm0.13$ \\
79666497970013398 & 268.3915 & 65.8653 & 0.10 & 23.1 & 1.7 & $8.35\pm0.07$ & $-1.75\pm0.26$ & $7.73\pm0.86$ & $0.08\pm0.14$ \\
79666497970008145 & 268.3952 & 65.8180 & 0.08 & 21.8 & 2.6 & $9.39\pm0.07$ & $-1.16\pm0.21$ & $8.77\pm0.21$ & $0.21\pm0.10$ \\
79666497969981859 & 268.3969 & 65.7910 & 0.07 & 24.2 & 1.0 & $7.82\pm0.09$ & $-2.85\pm0.89$ & $7.22\pm0.96$ & $0.19\pm0.29$ \\
79666497970012627 & 268.4002 & 65.8522 & 0.06 & 22.2 & 2.0 & $8.46\pm0.04$ & $-1.21\pm0.06$ & $7.23\pm0.58$ & $0.01\pm0.01$ \\
79666493675038361 & 268.4012 & 65.6223 & 0.07 & 20.7 & 2.7 & $10.11\pm0.02$ & $-4.77\pm5.24$ & $8.97\pm0.03$ & $0.28\pm0.02$ \\
79666493675030336 & 268.4038 & 65.5510 & 0.09 & 23.2 & 2.4 & $8.54\pm0.06$ & $-1.86\pm0.28$ & $7.18\pm0.76$ & $0.02\pm0.04$ \\
79666497970013083 & 268.4054 & 65.8565 & 0.10 & 24.0 & 2.1 & $8.25\pm0.08$ & $-2.29\pm0.63$ & $7.58\pm0.96$ & $0.13\pm0.21$ \\
79666497970002344 & 268.4055 & 65.7580 & 0.07 & 23.0 & 1.5 & $8.21\pm0.06$ & $-2.27\pm0.37$ & $6.95\pm0.88$ & $0.03\pm0.05$ \\
79666497970004416 & 268.4094 & 65.8101 & 0.04 & 20.3 & 2.0 & $9.71\pm0.05$ & $-0.48\pm0.03$ & $9.61\pm0.02$ & $0.62\pm0.01$ \\
79666497970003889 & 268.4139 & 65.7772 & 0.08 & 24.6 & 1.8 & $7.91\pm0.10$ & $-3.52\pm1.52$ & $6.77\pm1.07$ & $0.08\pm0.14$ \\
79666497970003283 & 268.4162 & 65.7708 & 0.09 & 23.7 & 1.5 & $8.32\pm0.08$ & $-2.14\pm0.54$ & $7.75\pm0.85$ & $0.17\pm0.25$ \\
79666493675046343 & 268.4240 & 65.6899 & 0.09 & 23.6 & 2.4 & $8.71\pm0.07$ & $-3.39\pm2.07$ & $7.45\pm0.85$ & $0.12\pm0.19$ \\
79666510854904782 & 268.4243 & 66.3879 & 0.09 & 23.5 & 2.6 & $8.55\pm0.07$ & $-1.66\pm0.31$ & $8.26\pm0.43$ & $0.31\pm0.25$ \\
79666493675037653 & 268.4270 & 65.6091 & 0.06 & 22.3 & 1.7 & $8.87\pm0.08$ & $-2.73\pm0.78$ & $7.31\pm0.65$ & $0.05\pm0.07$ \\
79666497969999011 & 268.4406 & 65.7340 & 0.08 & 22.4 & 1.5 & $8.62\pm0.07$ & $-1.76\pm0.33$ & $8.06\pm0.46$ & $0.17\pm0.15$ \\
79666519444847384 & 268.4414 & 66.8062 & 0.06 & 23.8 & 2.2 & $7.78\pm0.06$ & $-1.96\pm0.09$ & $7.06\pm0.69$ & $0.04\pm0.06$ \\
79666493675044592 & 268.4481 & 65.6755 & 0.09 & 24.1 & 1.8 & $7.89\pm0.06$ & $-2.27\pm0.30$ & $7.14\pm0.99$ & $0.06\pm0.11$ \\
79666506559949118 & 268.4485 & 66.2422 & 0.08 & 24.0 & 2.1 & $8.27\pm0.08$ & $-2.29\pm0.39$ & $7.26\pm0.95$ & $0.06\pm0.11$ \\
80093512208509467 & 268.4581 & 66.9450 & 0.08 & 23.5 & 2.4 & $8.82\pm0.07$ & $-2.06\pm0.56$ & $8.09\pm0.51$ & $0.25\pm0.24$ \\
79666506559931642 & 268.4666 & 66.1124 & $0.04^{sp}$ & 20.3 & 1.9 & $9.63\pm0.03$ & $-3.42\pm0.97$ & $7.13\pm0.53$ & $0.01\pm0.01$ \\
79666360531054344 & 268.4716 & 65.8425 & 0.05 & 21.7 & 3.9 & $9.25\pm0.05$ & $-1.09\pm0.16$ & $8.82\pm0.04$ & $0.24\pm0.04$ \\
79666360531053541 & 268.4723 & 65.8104 & 0.08 & 21.9 & 2.1 & $9.34\pm0.06$ & $-1.04\pm0.13$ & $9.23\pm0.03$ & $0.73\pm0.06$ \\
\hline
\end{tabular}
\tablefoot{(1) HSC ID of the source \citep{oi2021}; (2-3) Sky coordinates of the source based on the HSC detection; (4) Redshift from \citet{huang2021}. Galaxies with a spectroscopic redshift are marked with a superscript 'sp'; (5-6) Average \textit{r}-band surface brightness within the effective radius (in units of \magperarcsec{}) and the effective radius obtained from radial profile fitting; (7-10) Stellar mass, star formation rate, total infrared luminosity, and dust attenuation in the \textit{V}-band, respectively, from \cigale{} SED fitting. The error bars are the $1\sigma$ uncertainties from the Bayesian analysis of \cigale{}.}
\end{table*}

\end{appendix}

\end{document}